\documentclass[12pt,article]{elsarticle}
\usepackage{lineno, hyperref}
\usepackage{graphicx}
\usepackage{adjustbox}
\usepackage{graphics,bm,graphicx,psfrag,amssymb,amsmath,rotating,color}
\usepackage{mathtools}
\usepackage{mathrsfs}
\usepackage{setspace}
\usepackage{filecontents}
\usepackage{caption}
\usepackage{setspace, caption}
\usepackage{lipsum}
\usepackage{booktabs}
\usepackage[titletoc, title]{appendix}
\usepackage[utf8]{inputenc}
\usepackage{float}
\usepackage[a4paper,bindingoffset=0.2in,%
            left=1in,right=1in,top=1in,bottom=1in,%
            footskip=.25in]{geometry}
\usepackage{tablefootnote}
\usepackage{threeparttable}
\usepackage{gensymb}            
\begin{document}

\begin{frontmatter}

\title{Selection of axial dipole from a seed magnetic field in 
rapidly rotating dynamo models}

\author{Subhajit Kar}




\begin{abstract}
In this study, we investigate preferences of dipolar magnetic structure from a seed magnetic field in the rapidly 
rotating spherical shell dynamo models. In this study, we set up a realistic model to show the effect of the Lorentz 
force in the polarity selection. The important results that has come out from our study is that 
the magnetic field acts on the flow much before the saturation. Our study suggests that the growth of the magnetic field is not a kinematic 
effect as one might think off, rather a dynamic effect. This dynamic effect grows as the field generated with time and finally brings the saturation 
to the dynamo action. Previous studies show that Lorentz force effect the flow when Elsasser number more or less 1 and the studies were focused on 
the saturation by looking at the time-averaged quantities. However, in this study, we show a clear effect of the Lorentz force even at Elsasser number 
of $0.3-0.4$. To show the effect of the Lorentz force, we did two different simulations, one is a nonlinear model and another is kinematic model 
and shows that how a magnetic field can change the flow structure and by doing that the 
generated field changes, while this kind of behavior is not observed in kinematic  dynamo models. This study shows a scale dependent 
behaviour of the kinetic helicity at two different spectral range. 
\end{abstract}

\begin{keyword}
\texttt{elsarticle.cls}\sep \LaTeX\sep Elsevier \sep template
\MSC[2010] 00-01\sep  99-00
\end{keyword}

\end{frontmatter}


\section{Introduction}
The generation of magnetic field inside a rapidly rotating planetary core 
is a fundamental problem in geophysics. 
There has been considerable progress in modelling convection in rapidly 
rotating spherical shells, and the dynamo action; see for example the  
recent reviews \cite{Kono2002} , \cite{Roberts2013}. 
It is widely believed that the fluid in planetary cores is stirred by the 
columnar convection under rapid rotation which transports the heat from deep core 
to outer boundary, though precessional effects may also be involved in magnetic 
field production.
It is suggested that magnetic field of earth is maintained by thermal 
as well as compositional convection though latter is most dominant. 
However, the convection driven dynamos shows different types magnetic field 
structure can be generated - axial dipole, quadrupole \cite{Busse2006} \cite{Grote1999} 
\cite{Grote2000}, or even equatorial dipole \cite{Ishihara2002} \cite{Aubert2004}. 
However, axial dipole structure is very commom in rapidly rotating dynamos, 
where the inertia (nonlinear part of the advection) in the equation of motion is very small. 
Estimates of the core flow velocity (see, \cite{Jones2007}) suggest that inertia is 
significant in planetary cores at length scales so small that magnetic diffusion is 
very rapid, so the expected regime is indeed where axial dipole dominance is found. 
This may explain why most planetary dynamos are approximately dipolar, 
the only exceptions being Uranus and Neptune, whose physical properties are 
poorly understood.
Though Earth lies in a rapidly rotating system.  
Magnetic field structure from observation at core-mantle boundary (CMB) suggests 
that it is dipole. 
Magnetic field structure from 
observation at core-mantle boundary (CMB) suggests that it is dipole. 
Most of the numerical simulations achieved dipole field as most dominant. 
But, the question arises why this structure is the most likely solution 
from spherical dynamo models. 
One way to study it is by having a flow field which can closely mimic the 
features (like - differntial rotation, meriodinal circulations) of a rotating systems 
and see whether the dipole solution is favorable than any other solutions. 
But this approach is a linear, where back reaction of the magnetic field on 
flow is neglected. 
But in a planetary core, the effect of back reaction is imporatant and 
therefore, study with a linear approach is incomplete. 
In this study we will show how back reaction can alter the growth of 
magnetic fieled and also its structure.    
\par 

In a recent stduy by Schaeffer et al. \cite{Schaffer2006}, the authors has used using quasi-geostrophic 
formulation where, $z$-vorticity equation is computed at equatorial plane and extrapolated to the 
full sphere whereas, induction equation is solved in full sphere. 
Since, it is a quasi-geostrophic model, so $u_s$ and $u_{\phi}$ are constant 
along the direction of rotation (independent of $z$). 
This dynamo model excited stewartson layer by rotating the polar cap 
(the imaginary cylinder of radius of inner core touches the outer boundary) 
at different rotation than the outside. 
So, it will generate a geostrophic shear layer adjacent to the tangent cylinder. 
Certainly, this low $Pm$ models will have different characteristics than the 
usual convective dynamo models where geostrophic has to break down by convection 
itself though it will maintain the same velocity symmetry about the equator as 
dynamo models at close to onset of convection. 
The forcing of dynamo is defined by a non-dimensional quantity called Rossby number 
($Ro = \frac {\Delta \Omega}{\Omega}$), where $\Delta \Omega$ 
is the differential rotation rate of the inner core. 
Above a critical value of $Ro$ the stewartson layer will become unstable and 
generate a Rossby waves.
In this dynamo models both symmetry of magnetic fields exist 
(namely, dipole and quadrupole). 
The authors noted that in this configuration the quadrupole family has lesser 
critical magnetic Reynolds number than the dipole to excite. 
Though their energy spectrum suggested that toroidal magnetic energy
is more dominant than poloidal magnetic energy (about four orders 
of magnitude difference). 
It is an $\alpha$-$\omega$ dynamo model, where the shear inside the stewartson layer 
will produced toroidal field from poloidal component, which is famously known as 
$\omega$ effect whereas, the vortices produced by the Rossby waves will help to 
generate the poloidal field from the toroidal component (known as $\alpha$ effect) 
and this is how the dynamo regeneration cycle will be completed. 
But a steady flow fails to produce any dynamo (kinematic) actions because in this 
scenario there is no Rossby waves and hence, the dynamo regeneration cycle will be stopped.

In a study by Aubert et al.\cite{Aubert2004}, the authors has shown a self-consistent numerical 
dynamo models where both axial and equatorial dipoles can exist in a parameters regime of 
intermediate shell thickness (thickness ratio $\geq 0.5$) and close to onset of convection. 
Their numerical simulations suggest that at close to onset of convection, equatorial dipole is 
favored whereas at slightly supercritical convection the field is dominated axial dipole. 
Also they have shown that once the strong axial dipole is setup, it is very difficult to kill 
it off by decreasing the Rayleigh number, which is known 
as subcritical behaviour of dynamo. Most of their runs are at shell thickness ratio of greater 
than 0.5. They found that the equatorial dynamo solutions exists only at intermediate
shell thickness ratio (~0.45 - 0.7) but their strength is very low compare to the axial dipole. 

Generation of magnetic field by convective dynamo models in a spherical shell were
studied in case of Prandtl number of order one and a broad range of magnetic	
Prandtl number\cite{Grote1999}. The authors find regular and chaotic dipolar dynamos, 
quadrupolar dynamos and hemispherical dynamos at different range of magnetic Prandtl number. 
The basic state temperature contains an uniform heat source and velocity boundary 
condition is stress-free on both sides of the spherical shell. 
The non-dimensional parameters are - $Ek$ = $10^{-4}$, $Pr=1$. 
They found that quadrupolar dynamos are preferred at $Pm$ of order one while 
hemispherical dynamos are most at $Pm$ of order 10. 


Kinematic models help to find out what kind of flow structures can allow to have 
a dynamo action. Several people have tried to demonstrate the dynamo action with
simplified flow structure which can mimic the flow of a rotating sphere(main 
features are - differential rotation, meridional circulations etc). 
One of most cherished flow structure is Kumar-Roberts flow (in short KR flow). 
This flow contains differential rotation ($T_{1}^{0}$), meridional circulation 
($P_{2}^{0}$) and two convective rolls ($P_{2}^{2c}$ and $P_{2}^{2s}$). 
In this paper, the authors explain the dynamo action by varying the above flow 
features by solving induction equation as eigenvalue problem where, magnetic 
Reynolds number ($Rm$) is taken as eigenvalue. 
Lower value of $Rm$ represents the easy of dynamo action. 
To get the confidence of their numerical convergence, the authors tried the solution 
near to Braginsky limit ($Rm$ $\rightarrow$ $\infty$) by having strong axisymmetric 
differential rotation and small amount of meriodinal circulation and they found 
that differential rotation promotes axisymmetric toroidal field, whereas meridional 
circulation helps to generate axisymmetric poloidal field. 
They also found that non-axisymmetric helicity helps dynamo action by reducing 
critical $Rm$.



\section{Numerical method}
We consider an electrically conducting fluid confined in a spherical shell 
rotating with a constant angular velocity around $z$-axis. The radius of 
inner and outer spherical surfaces are $r_i$, $r_o$ respectively, and ratio
is chosen to be Earth like, 0.35. The governing equations considered are 
in the Boussinesq approximation. Lengths are scaled by thickness of the 
spherical shell $L$, and time is scaled by the magnetic diffusion time ($t_d$),
($L^2/\eta$), where, $\eta$ is magnetic diffusivity. The velocity field $\bm{u}$
is scaled by $\eta/L$, and the magnetic field $\bm{B}$ is scaled by 
$(2 \Omega \rho \mu \eta)^{1/2}$, where $\Omega$ is the rotation rate, 
$\rho$ is the fluid density and $\mu$ is the free space magnetic permeability.
\par The non-dimensional magnetohydrodynamic(MHD) eqautions for velocity, 
temperature and magnetic fields are
\begin{equation}\label{eq1}
\begin{aligned}
{E}{Pm}^{-1} \biggl(\frac {\partial \bm{u}}{\partial t} + 
(\nabla \times \bm{u}) \times \bm{u}  \biggr) + 
\hat{\bm{z}} \times \bm{u} = -\nabla p^*  + {q} {Ra} T \bm{r} \\
+ (\nabla \times \bm{B}) \times \bm{B} + {E} \nabla^2 \bm{u}
\end{aligned}
\end{equation} 
\begin{equation}\label{eq2}
\begin{aligned}
\frac {\partial T}{\partial t} + \bm{u} \cdot \nabla T = {q} \nabla^2 T
\end{aligned}
\end{equation}
\begin{equation}\label{eq3}
\begin{aligned}
\frac {\partial \bm{B}}{\partial t} = \nabla \times (\bm{u} \times \bm{B}) 
+  \nabla^2 \bm{B}
\end{aligned}
\end{equation}
\begin{equation}\label{eq4}
\begin{aligned}
\nabla \cdot \bm{u} = \nabla \cdot \bm{B} = 0
\end{aligned}
\end{equation}

The $\nabla p^*$ in Eq. \ref{eq1} is  a modified pressure given by  
$p + \frac{1}{2} E Pm^{-1} | \bm{u}|^2 $, where $p$ is the  fluid pressure. 
The dimensionless parameters in Eqs. \ref{eq1} - \ref{eq3} are the Ekman number, 
$E$ = ${\nu}/{2\Omega L^2}$, that measures the ratio of viscous to rotational forces,
the Prandtl number, $Pr$=${\nu}/{\kappa}$ that gives the ratio of viscous to thermal
diffusivities, the magnetic Prandtl number, $Pm$=${\nu}/{\eta}$ that gives the ratio 
of viscous to magnetic diffusivities and the 'modified' Rayleigh number 
(product of classical Rayleigh number and Ekman number) is given by 
${g_{o} \alpha \Delta T L}/{2 \Omega \kappa}$,
where $g_o$ is the gravitational acceleration acting in radially inward, $\alpha$
is the coefficient of thermal expansion, $\Delta T$ is the superadiabatic
temperature difference between boundaries and $\kappa$ is the thermal diffusivity. 
The basic-state non-dimensional temperature distribution is a conventional basal 
heating $T_o(r)$ = $\beta/r$, where $\beta={r_i}{r_o}$. The velocity, temperature and 
magnetic fields satisfy the no-slip, isothermal and electrically insulating 
conditions, respectively. In order to characterize the results of the simulations,
kinetic energy ($E_k$), magnetic energy $(E_m)$ and Elsasser number $(\Lambda)$ are, 

\begin{equation}\label{eq5}
\begin{aligned}
E_k=\frac{1}{2} \int_{V} |\bm{u}|^2 dV, \hspace{0.3in} 
E_m=\frac{Pm}{2E} \int_{V} |\bm{B}|^2 dV, \hspace{0.3in} 
\Lambda=\sqrt{\frac{2{Pm}{E_m}}{{E}{V}}}.
\end{aligned}
\end{equation}
where, $V$ represents volume of the spherical shell. 

The non-dimensional equations are solved using a pseudospectral method in which 
the velocity and magnetic field are expanded as toroidal and poloidal vectors, 
for example the toroidal-poloidal decomposition of velocity field is written as 
\cite{Bullard1954}
\begin{equation}\label{eq6}
\begin{aligned}
\bm{u} = \sum\limits_{m=0}^{M} \sum\limits_{l=1}^{L} (\bm{u}_{lm}^T + \bm{u}_{lm}^P) 
\end{aligned}
\end{equation}
where, superscripts $T$ and $P$ denotes toroidal and poloidal parts of $\bm{u}$ at a given
degree $l$ and order $m$, respectively. Similar way we can expand magnetic field $(\bm{B})$.
\subsection{Equation of magnetic energy}    
The induction term of Eq. \ref{eq2} can be rewritten in terms of toroidal and 
poloidal forms of velocity and magnetic field -
\begin{equation}\label{eq7}
\begin{aligned}
\frac {\partial \bm{B}}{\partial t} 
= \biggl( \nabla \times (\bm{u}^P \times \bm{B}^P) 
+ \nabla \times (\bm{u}^P \times \bm{B}^T)  +  
\nabla \times (\bm{u}^T \times \bm{B}^P)  + 
\\ \nabla \times (\bm{u}^T \times \bm{B}^T)   \biggr)   +  \nabla^2 \bm{B}
\end{aligned}
\end{equation}
The magnetic energy equation integrating over spherical shell volume ($V$)
\begin{equation}\label{eq8}
\begin{aligned}
\frac{Pm}{2 E}\frac {d} {dt} \int_{V} |\bm{B}|^2 dV = 
\frac{Pm}{E} \int_{V} \bm{B} \cdotp \biggl(\nabla \times (\bm{u}^P \times \bm{B}^P)
+ \nabla \times (\bm{u}^P \times \bm{B}^T) +  
\\ \nabla \times (\bm{u}^T \times \bm{B}^P) + 
\nabla \times (\bm{u}^T \times \bm{B}^T) \biggr) dV 
+ \frac{Pm}{E} \int_{V} \bm{B} \cdotp \nabla^2 \bm{B} dV
\end{aligned}
\end{equation}

Therefore, the energy equation for $\bm{B}_{20}^T$ becomes 
\begin{equation}\label{eq9}
\begin{aligned}
\frac{Pm}{2 E} \frac {d} {dt} \int_{V} |\bm{B}_{20}^T|^2 dV = 
\frac{Pm}{E} \int_{V} \bm{B}_{20}^T \cdotp \biggl(\nabla \times 
(\bm{u}^P \times \bm{B}^P)
+ \nabla \times
(\bm{u}^P \times \bm{B}^T) + \\ 
\nabla \times (\bm{u}^T \times \bm{B}^P) + 
\nabla \times (\bm{u}^T \times \bm{B}^T) \biggr) dV + 
\frac{Pm}{E} \int_{V} \bm{B}_{20}^T \cdotp \nabla^2 \bm{B}^T_{20} dV
\end{aligned}
\end{equation}

Since, Bullard's selection rule \cite{Bullard1954} (p 229) suggests $\bm{u}^T$ and $\bm{B}^T$ 
can not generate $\bm{B}^P$. Hence, the energy equation for $\bm{B}_{10}^P$ becomes
\begin{equation}\label{eq10}
\begin{aligned}
\frac{Pm}{2 E} \frac {d} {dt} \int_{V} |\bm{B}_{10}^P|^2 dV
= \frac{Pm}{E} \int_{V} \bm{B}_{10}^P \cdotp \biggl(\nabla \times (\bm{u}^P \times \bm{B}^P)
+  \nabla \times (\bm{u}^P \times \bm{B}^T) + \\
\nabla \times (\bm{u}^T \times \bm{B}^P) \biggr) dV + 
\frac{Pm}{E} \int_{V} \bm{B}_{10}^P \cdotp \nabla^2 \bm{B}^P_{10} dV
\end{aligned}
\end{equation}
\section{Results}
\noindent The simulations has two steps. First, a pure hydrodynamic simulation is performed at 
each $Ra$ number to obtain saturated state. Then, a seed magnetic field 
 of $\Lambda = 0.01$ is used in the saturated hydrodynamic state to start both 
kinematic and nonlinear dynamo simulations \cite{Ishihara2002}. 
For kinematic simulations, Lorentz force term was dropped from mometum equation and 
start time-stepping all three equations \cite{Aubert2004}. For a frozen flow (saturated
velocity field at a given time) we do not find any dynamo action in
our low-E parameter regimes as classical kinematic theory deals with \cite{Schaffer2006}.
This suggests that Rossby wave plays an important role in magnetic field generation
in the rapidly rotating spherical shell dynamo models \cite{Avalos2009}. 
The roles of drifting frequency of
the columnar structure (for exmaple, Kumar-Roberts flow) on dynamo action is briefly studied 
in the framework of kinematic models \cite{Willis2004}. The authors shown that for a given
manetic magnetic Reynolds number ($Rm$) the growth rate of dynamo increases
with drifting frequency of the columns in wide range of the parameter space. 
Though in our study structure of the initial velocity field is far more 
complex. The focus of this study is to quantify the roles of rapid rotation and manetic 
field in polarity selection from a seed magnetic field  ($\Lambda=0.01$) in the 
nonlinear and kinematic regimes.
We perform calculations at two different Ekman numbers. 
To keep the effect of interia low on the simulations \cite{Sreeni2006}, 
we fix our parameters $E=1.2 \times 10^{-5}, Pr=Pm=5$ and $E=1.2 \times 10^{-6}, Pr=Pm=1$.  
Table \ref{tab:example_tab} summarises the details of simulations and computed value of
Elasser number at saturated nonlinear dynamos. For comparison purposes kinematic simulations
are performed in each case. 

\subsection{Role of symmetry in polarity selection}
\begin{figure}[t!]
	\centering
	\includegraphics[width=\textwidth]{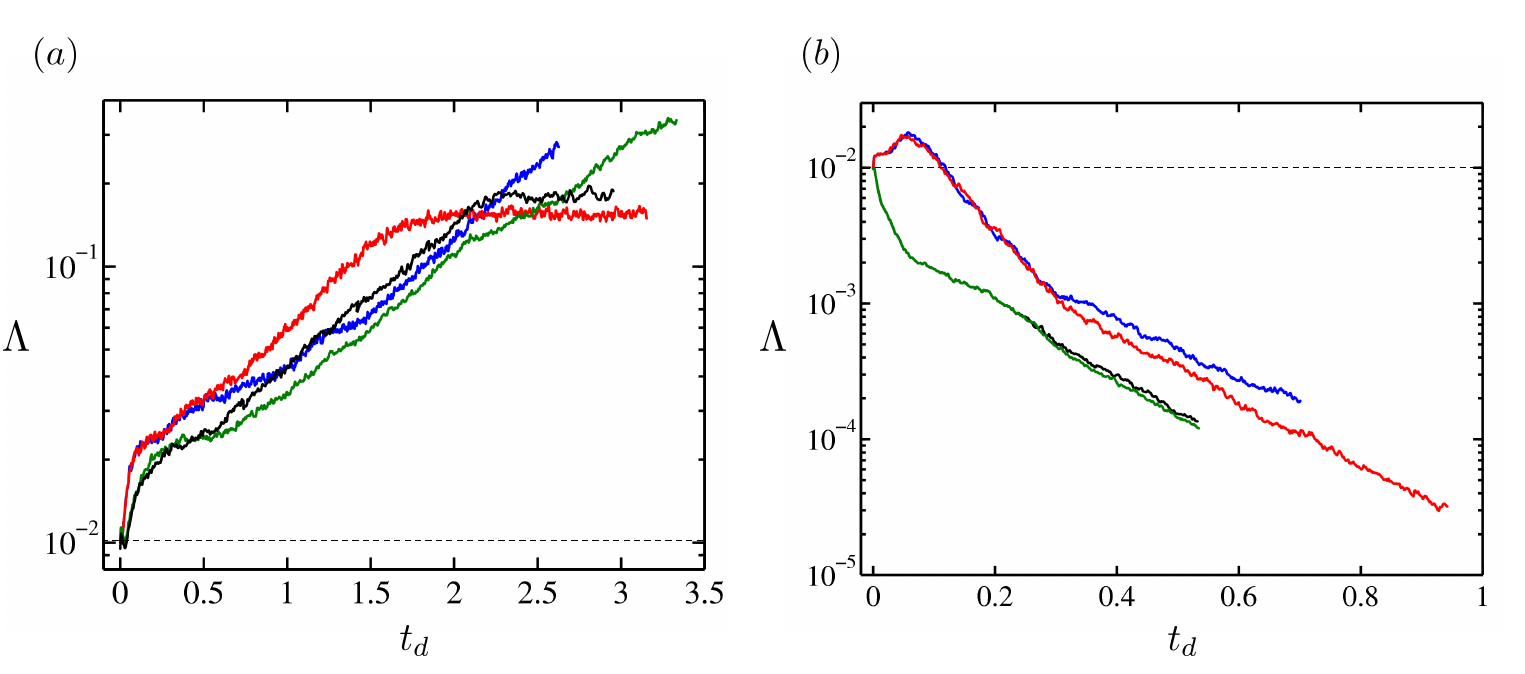}
	\caption{Evolution of Elsasser number with magnetic diffusion time for 
	dynamo simulations at $E=1.2 \times 10^{-5}, Pr=Pm=5, Ra=140$. 
	The structure of  starting seed magnetic field are - 
	$(a)$ $\bm{B}_{10}^P$ (red and blue), $\bm{B}_{30}^P$ (black and green) 
	$(b)$ $\bm{B}_{30}^P$ (red and blue), $\bm{B}_{31}^P$ (black and green).   
	Red and black lines respresent nonlinear and black and green lines for 
	kinematic cases.}
	\label{fig:fig1}
\end{figure}

\noindent 
Evolution of Elsasser numbers for different symmetries belong to dipole and 
quadrupole families for simulations at $E=1.2 \times 10^{-5}, Pr=Pm=5$, $Ra=140$ 
are shown in Fig. \ref{fig:fig1}.
For low-$Ra$ case, the symmetry of velocity field is preserved under rapid rotation i.e., 
the flow is symmetric about the equator. Such kind of flow preserve the
equatorial symmetric or antisymmetric of the initial magnetic field throughout 
the simulations in a kinematic dynamo model.
In this case we can see from Fig. \ref{fig:fig1} that the initial magnetic field 
belong to dipole family is growing while quadrupole symmetries are falling in 
both nonlinear and kinematic cases. \par
{\renewcommand{\arraystretch}{1.5} 
\begin{table}[t!]
\begin{center}
  \caption{Parameters used in kinematic and nonlinear dynamo simulations and computed 
  $E_m/E_k$, $\Lambda$, $\Lambda_{AD}$ for saturated nonlinear simulations.
  $\Lambda_{AD}$ denotes Elsasser number based on
  magnetic energy contain in $\bm{B}^P_{10}$ and $\bm{B}^T_{20}$. 
  D corresponds to dipoe and ND denotes nondipole magnetic field
  structure at the outer boundary of the spherical shell.
  For each $Ra$, a hydrodynamic simulation was performed to use the velocity field in the 
  dynamo model. 
  $\bm{B}_{init}$ represents the starting magnetic field structure. 
  For $Ra=140$, symmetry of the velocity field is
  broken by sowing antisymmetric energy 1\% of total kinetic energy.}
\begin{threeparttable}
\scalebox{1.08}{
  \begin{tabular}{c c c c c c c c c}
     \hline 
 $E$ & $Ra$ & $Pr=Pm$ & $l_{max}$ & $N_r$ 
 & $\bm{B}_{init}$  & $\Lambda$ & $\Lambda_{AD}$ & $\bm{B}_{r=r_o}$  \\
     \hline 
$1.2 \times 10^{-5}$  & $140$ & $5$  & $80$ & $96$ 
& $\bm{B}^P_{10}$ & $0.11$ & $0.09$ & D(D) \\

$1.2 \times 10^{-5}$  & $140$ & $5$  & $80$ & $96$ 
& $\bm{B}^P_{20}$ & no dynamo & $-$ & $-$ \\

$1.2 \times 10^{-5}$  & $140$ & $5$  & $80$ & $96$ 
& ${\bm{B}^P_{20}}^{\tnote{1}}$ & $0.11$ 
& $0.09^{\tnote{2}}$ & D(D) \\

$1.2 \times 10^{-5}$  & $140$ & $5$  & $80$ & $96$ 
& $\bm{B}^P_{30}$ & $0.11$ & $0.09$ & D(D) \\

$1.2 \times 10^{-5}$  & $140$ & $5$  & $80$ & $96$ 
& $\bm{B}^P_{31}$ & no dynamo & $-$ & $-$ \\

$1.2 \times 10^{-5}$  & $140$ & $5$  & $80$ & $96$ 
& mixed${\tnote{3}}$  & $0.11$ & $0.09$ & D(D) \\

$1.2 \times 10^{-5}$  & $220$ & $5$  & $100$ & $120$ 
& $\bm{B}^P_{10}$ & $0.99$ & $0.37$  & D(ND) \\

$1.2 \times 10^{-5}$  & $220$ & $5$  & $100$ & $120$ 
& ${\bm{B}^P_{20}}$ & $0.99$ & $0.37^{\tnote{2}}$  & D(ND) \\

$1.2 \times 10^{-6}$  & $200$ & $1$  & $180$ & $220$ 
& $\bm{B}^P_{10}$ & $0.1$ & $0.08$  & D(D) \\

$1.2 \times 10^{-6}$  & $400$ & $1$  & $220$ & $220$ 
& $\bm{B}^P_{10}$ & $0.98$ & $0.35$ & D(ND) \\

$1.2 \times 10^{-6}$  & $400$ & $1$  & $220$ & $220$ 
& ${\bm{B}^P_{20}}$ & $0.98$ & $0.35^{\tnote{2}}$ & D(ND) \\

$3 \times 10^{-7}$  & $540$ & $1$  & $320$ & $360$ 
& ${\bm{B}^P_{20}}$ & $1.56$ & $0.35$ & D(ND) \\
     \hline
  \end{tabular}
}
\begin{tablenotes}
\item[$1$] artifically break equatorial symmetry of the impose initial 
velocity field.
\item[$2$] $\bm{B}^P_{20}$ convert to $\bm{B}^P_{10}$.
\item[$3$] 99.99\% $\Lambda_{20}^P$ + 0.01\% $\Lambda_{10}^P$. 
\end{tablenotes}
\end{threeparttable}
  \label{tab:example_tab}
\end{center}
\end{table} 
}

\begin{figure}[t!]
	\centering
	\includegraphics[width=\textwidth]{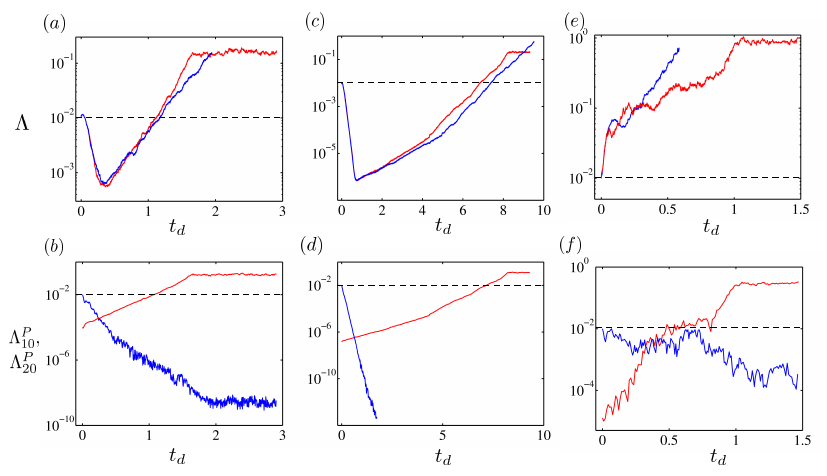}
	\caption{Evolution of Elsasser number of $\bm{B}_{10}^P$ (red line) and 
	 $\bm{B}_{20}^P$ (blue line) for the nonlinear dynamo simulations
	 for $E=1.2 \times 10^{-5}, Pr=Pm=5$. 
	 The structure of starting seed magnetic field are - 
	 $(a)$ mixed (99.99\% $\Lambda_{20}^P$ + 0.01\% $\Lambda_{10}^P$ ) and 
	 $(b)$ artificially broken velocity symmetry at $Ra=140$.
	 and $(c)$ $\bm{B}_{20}^P$ at $Ra=220$. }
	\label{fig:fig2}
\end{figure}

We choose to study further details about the role of symmetries of the 
velocity field and the magnetic field in polarity selection.
To keep the analysis simple, we study the growth/fall of the axial dipole and axial
quadrupole.  
Though, there is no dynamo action of pure axial dipole or axial quadrupole
as Cowling's antidynamo theroy suggests \cite{Cowling1934}. 
Nevertheless, we find that the growth of axial dipole is much higher than 
the any other field configuration in a successful dynamo. 
First, we consider an initial seed magnetic field ($\Lambda=0.01$) composed of 
99.99\% of $\bm{B}_{20}^P$ and 0.01\% of $\bm{B}_{10}^P$ and continue the simulations
in both kinematic and nonlinear models at $Ra=140$. 
Fig. \ref{fig:fig2} (a) illustrates that successful dynamos in both nonlinear
(red) and kinematic(blue) regimes unlike the case in Fig. \ref{fig:fig1} $(c)$.
Elsasser numbers of $\bm{B}_{10}^P$ and $\bm{B}_{20}^T$ for nonlinear dynamo cases 
in Fig. \ref{fig:fig2} ($b$) suggests that under preserved symmetry of velocity 
only axial dipole component of the field is growing and the axial quadrupole 
component is falling and saturated to a very small value due to back reaction effect. 
For kinematic case, we do expect the same behaviour, except
the saturation and the axial quadrupole will continue to fall. 
In second case, we break the symmetry of velocity field by sowing equatorial 
antisymmetric component of 1\% of total kinetic energy in case of 
initial magnetic field of $\bm{B}^P_{20}$ at same $Ra$. 
By doing so, we see that in Fig. \ref{fig:fig2} ($c$) that dynamo is growing 
and the harmonic analysis of energy of the nonlinear simulation in Fig. 
\ref{fig:fig2} ($d$) suggests that it is axial dipole component which is growing 
and axial quadrupole is falling. 
In high-$Ra$ case, where the symmetry of velocity field break down naturally under 
the rapid rotation, we see that in Fig. \ref{fig:fig2} ($e$ and $f$) that starting 
with $\bm{B}_{20}^P$ convert itself into dipole. 
This happen due to the fact that nonlinear product of equatorial antisymmetric 
of $\bm{u}$ and equatorial symmetric of $\bm{B}$ give rise to an antisymmetric 
$\bm{B}$ \cite{Gubbins1993}. 
The initial axial quadrupole (equatorial symmetric structure) interacts with the 
antisymmetric components of the flow and induced equatorial antisymmetric magnetic 
field, belong to dipole family. 
This happen in the very early phase of the simulation and this flow-field 
interaction is purely a kinematic effect. Therefore, we get a growing dynamo. 
\par 
We have tested the idea that for a rapidly rotating dynamo with equatorial 
symmetry broken flow always gives a successful dipolar dynamo by changing 
the initial structure of the magnetic field.    
In next paragraph, we explore the mechanisms to explain the failure of a 
quadrupolar dynamo and success of a dipolar dynamo. 
\begin{figure}[t!]
	\centering
	\includegraphics[width=\textwidth]{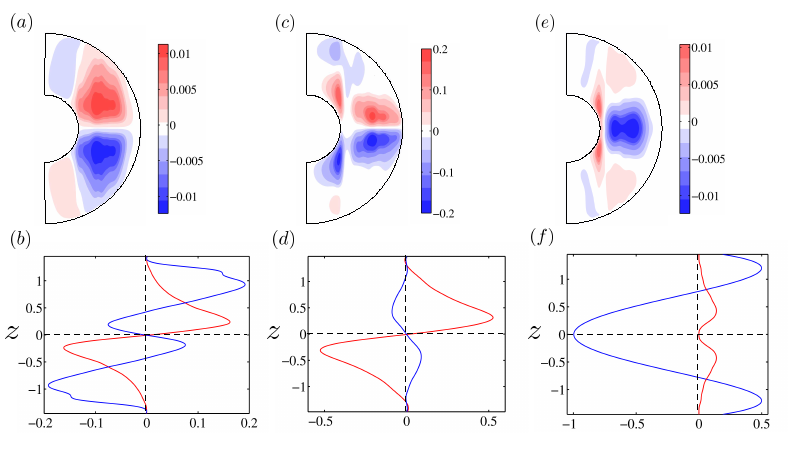}
	\caption{Snapshots of axisymmetric $B_\phi$ generation for kinematic dynamo 
	simulations at $E=1.2 \times 10^{-5}, Pr=Pm=5, Ra=140$. 
	The snapshots of first and third columns are at 
	$t_d$ = $3 \times 10^{-2}$ and the second column at $t_d$ = $1.5$. 
	Upper column shows snapshots of $B_\phi$ and bottom shows the generation terms. 
	Red and blue lines represent $s$ - $\phi$ averaged of 
	${B_z}{\partial u_\phi/\partial z}$ 
	and ${B_s}{\partial u_\phi/\partial s}$ respectively. }
	\label{fig:fig3}
\end{figure}
\par 
For simplicity, we use dynamo models starting from axial dipole
and axial quadrupole. 
To analysis the simulations, kinematic cases are used. 
Since, from Fig. \ref{fig:fig1}$ (c-d)$ we do not see much differences between 
nonlinear and kinematic for quadrupole run, while for dipole case, we see a 
growing dipolar dynamo in both cases.
Fig. \ref{fig:fig3} shows the axisymmetric $B_\phi$ generation for dipole and 
quadruple cases for kinematic dynamo simulations at $Ra=140$. 
For dipole case, initially the axisymmertic toroidal field is generated by both 
shearing of axial magnetic field ($B_z$) and radial magnetic field ($B_s$) in 
cylindrical coordinate by differential rotation, known
as $\omega$-effect \cite{Parker1955}. 
Further in time, we see it is the shearing of $B_z$ which supports the 
field generation. While in case of quadrupole, axisymmetric toroidal field is 
generated by a different mechanism from dipole and it is the radial magnetic field 
sheared by the zonal flow $(u_\phi)$ in cylindrical coordinate.
Time evolutions of Elsasser number for quadrupole case (Fig. \ref{fig:fig1} ($c$)), 
suggests that there is an initial peak for short period of time and 
then it is falling in time. 
This peak is attribute to the initial $B_\phi$ generation.
As we go for more complex field structure (for example, $\bm{B}^P_{31}$), 
we do not see any peak in magnetic energy and from begining of the simulation, 
the dynamo is falling.
For axial dipole case, $B_s$ is well above the equatorial plane.  
This suggests that the columnar structure of the flow helps to 
regenerate poloidal field from a toroidal field through a classical 
$\alpha$-effect \cite{Parker1955} at least in the kinematic regime 
and became a successful dynamo. 
Though, there is a difference in axial dipole
field generation between nonlinear and kinematic dynamo models. 
We will pursue this issue in coming subsection.
As in case of quadrupole, maximum value of $B_\phi$ is confined to the equator plane. 
Hence, the flow failed to generate any further poloidal magnetic field from the 
toroidal field through Parker classical $\alpha$-effect \cite{Parker1955}.
Therefore, the dynamo regeneration cycle (Toroidal $\leftrightarrow$ Poloidal) is 
stopped and become a failed dynamo. 
Nevertheless, it is clear the initial magnetic field structure 
determines fate of the dynamos in both kinematic and nonlinear case as 
long as rapid rotation maintains symmetry of the velocity field.
\subsection{Role of back reaction in polarity selection}
\noindent
In this study, we focus on generation of the dipole field.
Therefore, for simplicity we wish to restrict on the $Y^0_1$ of poloidal 
and $Y^0_2$ of toroidal magnetic field.
We separate these two harmonics from the induction equation and study their 
growth with time. 
\par
\begin{figure}[t!]
	\centering
	\includegraphics[width=1.0\textwidth]{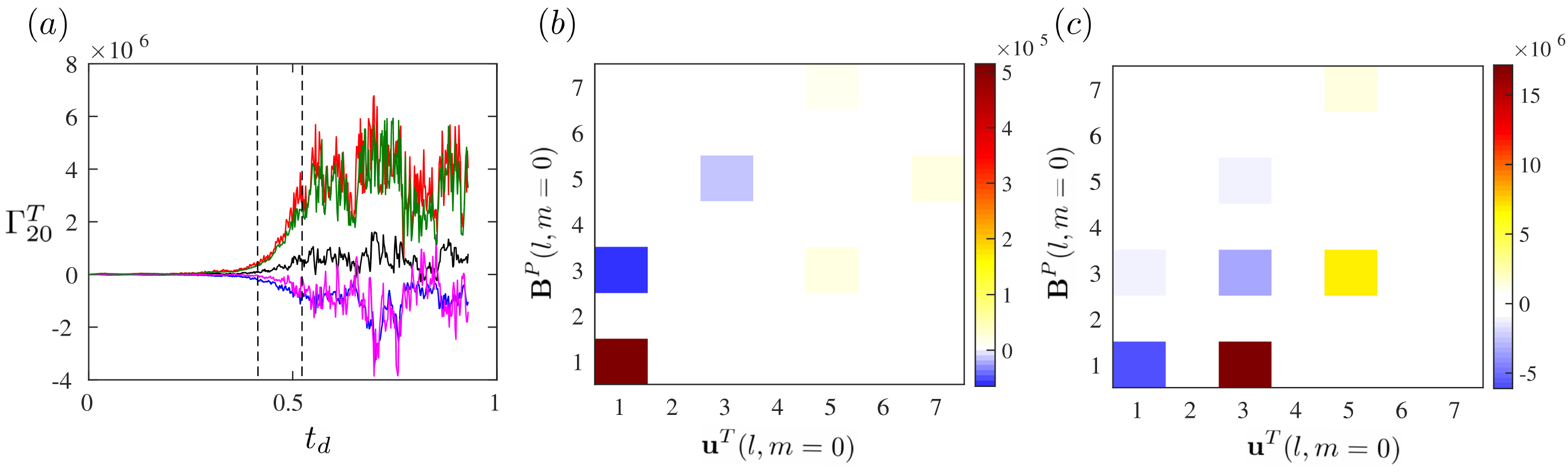}
	\caption{$\bm{B}_{20}^T$ generation mechanism in terms of energy for 
	$E=1.2 \times 10^{-5}, Ra=220$.  The top panel represents volume averaged of 
	induction term dotted with $\bm{B}_{20}^T$. The lines represent - 
	red - 
	$ (Pm/E) \bm{B}_{20}^T$.$\nabla \times (\bm{u}^T \times \bm{B}^P)$, 
	blue - 
	$ (Pm/E) \bm{B}_{20}^T$.$\nabla \times (\bm{u}^P \times \bm{B}^T)$, 
	black - 
	$ (Pm/E) \bm{B}_{20}^T$.$\nabla \times (\bm{u}^P \times \bm{B}^P)$, 
	magenta - 
	$ (Pm/E) \bm{B}_{20}^T$.$\nabla \times (\bm{u}^T \times \bm{B}^T)$ 
	and green - 
	$ (Pm/E) \bm{B}_{20}^T$.$\nabla \times (\bm{u}_{10}^T \times \bm{B}_{10}^T)$.
	The bottom panel represents energy matrix for 
	$(Pm/E) \bm{B}_{20}^T$.$\nabla \times (\bm{u}^T(m=0) \times \bm{B}^P(m=0))$
	at two different times. $(b)$ is at $t_d=0.414$ and $(c)$ is at $t_d=0.522$.}
	\label{fig:fig5}
\end{figure}

Fig. \ref{fig:fig5} shows the generation of $\bm{B}_{20}^T$ in nonlinear dynamo 
simulation for case of $E=1.2 \times 10^{-5}$, $Pr=Pm=5$, $Ra=220$ starting magnetic 
field as $\bm{B}^P_{10}$. 
Though as we see in the previous section that the starting 
magnetic strcuture does not play any role in the case of high-$Ra$.
Fig. \ref{fig:fig5} ($a$) shows the energy contribution to $\bm{B}_{20}^T$ through 
Eq. \ref{eq9}. 
The energy supplied to $\bm{B}_{20}^T$ is calculated using 
\begin{equation}\label{T1}
\begin{aligned}
\Gamma^{T}_{20} = \int_{V} \bm{B}^T_{20} \cdot \nabla \times 
(\bm{u} \times \bm{B})  dV
\end{aligned}
\end{equation}
and apply Bullard selection rule for different combinations of $\bm{u}$ and $\bm{B}$.
The maximum contribution is coming from induction term consists of
$\bm{u}^T$  and $\bm{B}^P$ (red line) and majorly it is axisymmetric structures 
(green line) that is contributing in the generation. 
Interestingly, there is no major role of back reaction of the magnetic field in 
changing the mechanism of generation, except saturation of the field. 
This suggests that the classical-$\omega$ effect is valid in nonlinear regime too 
as demonstrated previous \cite{Schrinner2012}. 
\begin{figure}[t!]
	\centering
	\includegraphics[width=0.6\textwidth]{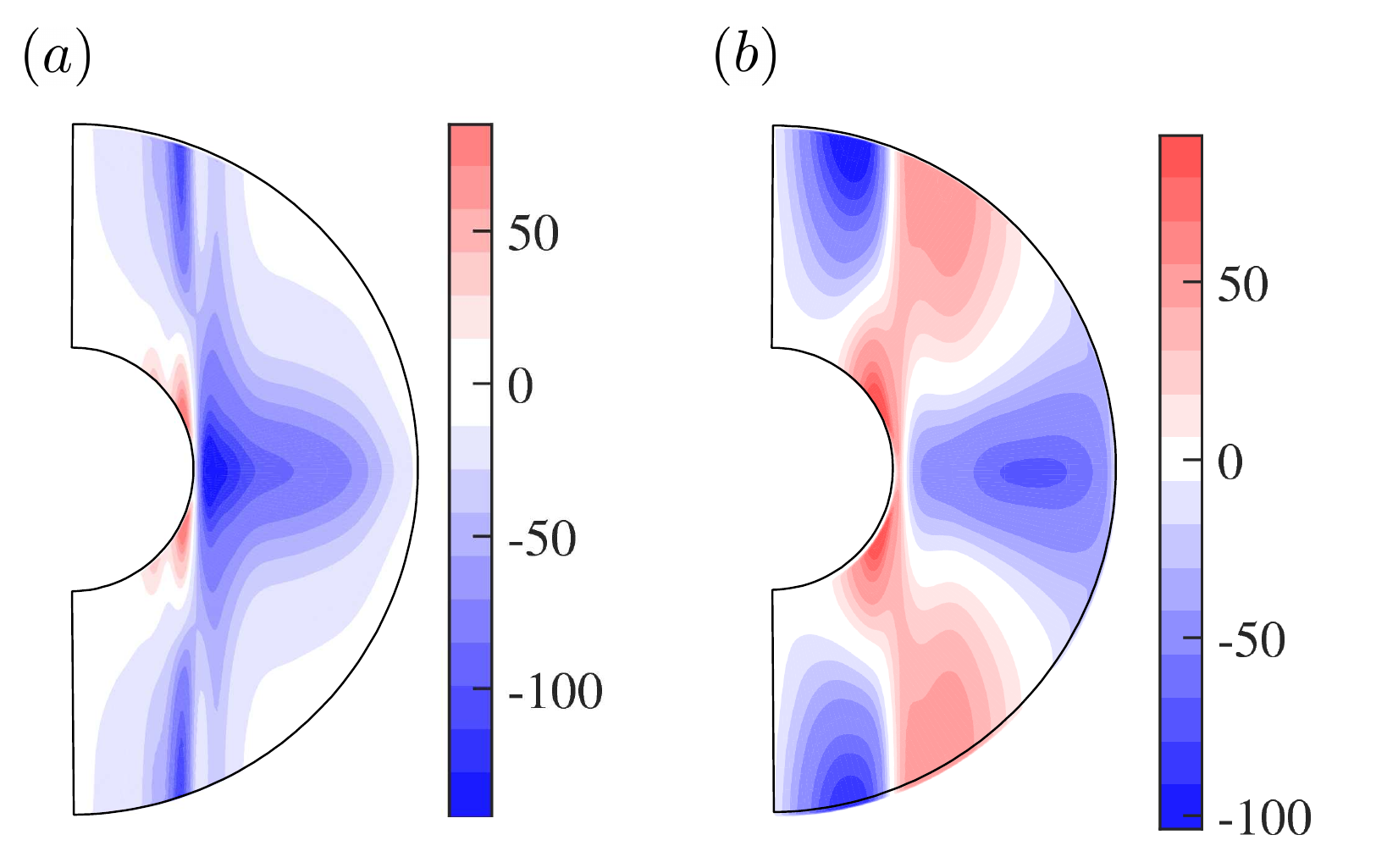}
	\caption{Meriodinal plots of time and azimuthally averaged zonal flow for 
	    simulations at $E=1.2 \times 10^{-5}$, $Ra=220$. $(a)$ represents
		non-magnetic and $(b)$ nonlinear dynamo in saturated phase.}
	\label{fig:fig6}
\end{figure}
As classical-$\omega$ effect suggests that it is the large scale flow
that supports the azimuthal field.
To verify this idea, we construct a energy matrix based on Bullard's selection rule 
\cite{Bullard1954} by keeping the structure of flow ($\bm{u}^T$) and field ($\bm{B}^P$) 
as axisymmetric ($m=0$).  
The bottom panel of Fig \ref{fig:fig5} shows the energy contribution of different 
degrees of the flow and field to the $\bm{B}_{20}^T$ at two different snapshot times
($t_d=0.414$ and $t_d=0.522$). 
Since our model start with a seed magnetic field, so we do expect the effect of back 
reaction on the flow after some time.
For Fig. ($b$), the energy matrix suggests that it is $Y^0_1$ of the toroidal flow 
that is helping in $\bm{B}_{20}^T$ generation. 
While for Fig. ($c$), it is $Y^0_1$ component of the toroidal flow supports the field. 
This clearly shows effect of back reaction on the large scale flow in time. 
This effect is attribute to effect of toroidal magnetic force on large scale zonal flow. 
In Fig. \ref{fig:fig6} shows the structure of time and azimuthally averaged zonal flow 
for non-magnetic ($a$) and nonlinear dynamo ($b$) models in the saturated state.
For non-magnetic case, structure of the zonal flow is geostrophic 
(dominantly $Y^0_1$), where as for nonlinear case, the dominant structure is mostly 
dominated by $Y^0_3$ \cite{Aubert2005}. 
Therefore, we see a close analogy between energy matrices and zonal flow structures.    
\par 
As we see in the previous paragraph that, axial dipole plays a major role in 
$\bm{B}_{20}^T$ generation.
Now, we look into mechanism of $\bm{B}_{10}^P$ generation. 
Here also case of starting magnetic field as axial dipole is considered. 
However we find that at low-$Ra$ regime,(for example, case at 
$E=1.2 \times 10^{-5}, Ra=140$) the poloidal magnetic field is dipolar structure in both 
kinematic and nonlinear dynamo models. 
However, differences in magnetic field structure between kinematic and nonlinear 
dynamos show up at high-$Ra$ $(Ra=220)$. 
\begin{figure}[t!]
	\centering
	\includegraphics[width=0.45\textwidth]{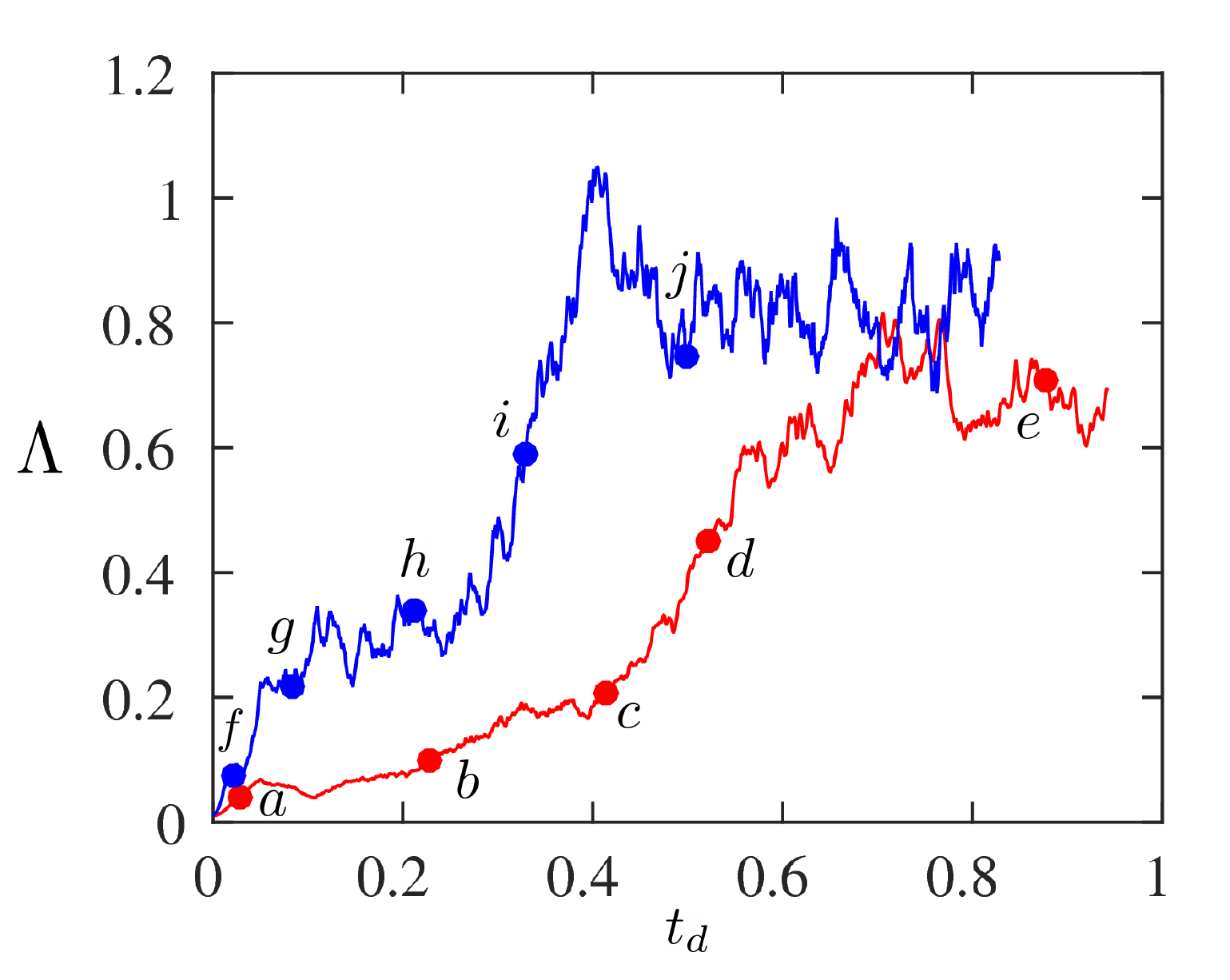}
	\caption{Evolution of Elsasser number for nonlinear dynamo simulations  
		at $E=1.2 \times 10^{-5}$, $Pr=Pm=5$ $Ra=220$ (red) and 
		$E=1.2 \times 10^{-6}$, $Pr=Pm=1$ $Ra=400$ (blue). 
		The alphabet numberings correspond to the
		snapshot times of Fig. \ref{fig:fig8}.
	 }
	\label{fig:fig7}
\end{figure}
\begin{figure}[t!]
	\centering
	\includegraphics[width=0.95\textwidth]{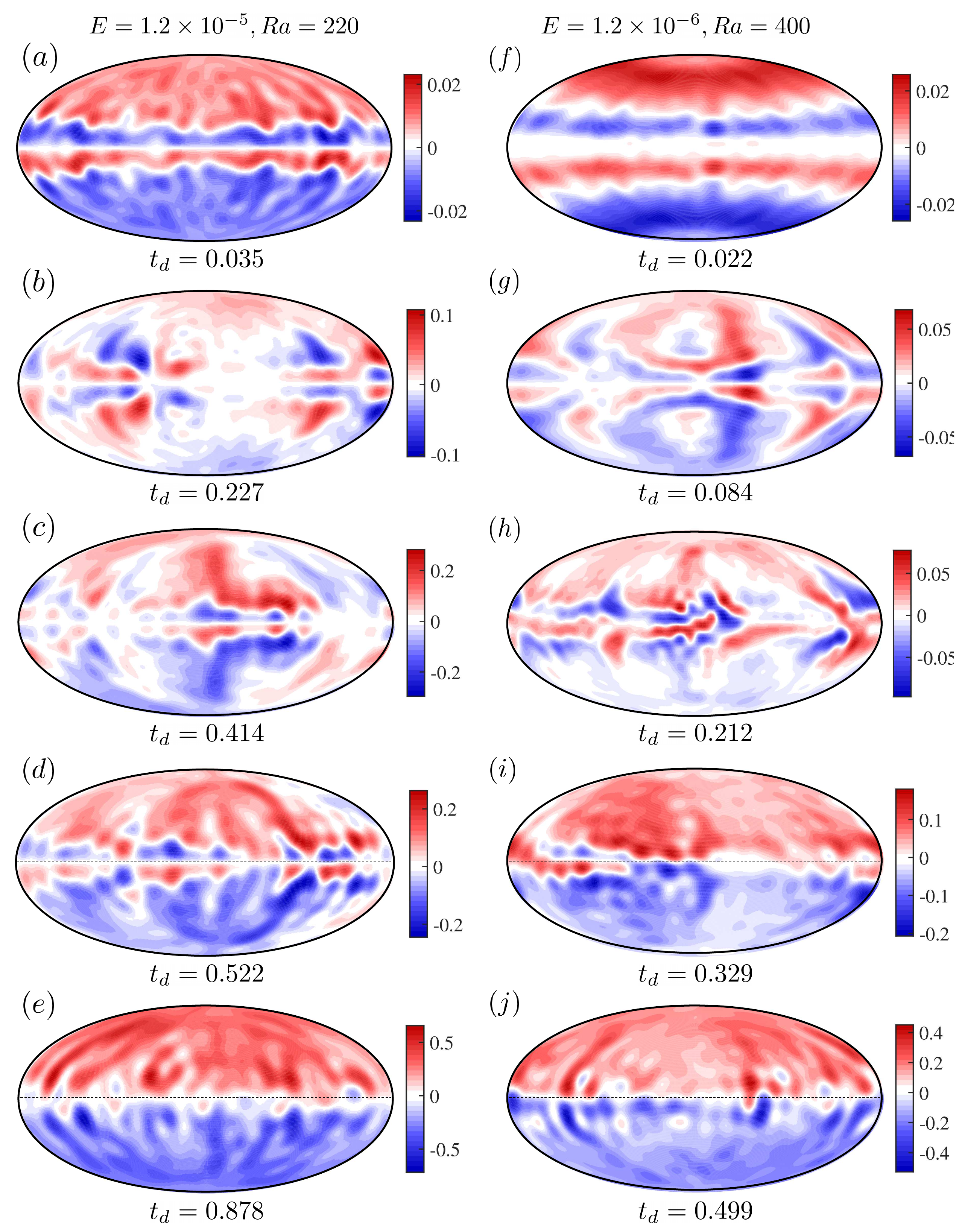}
	\caption{Evolution of radial magnetic field at the outer boundary for 
	nonlinear dynamo simulations. The model parameters are given at top of the figure. 
	Snapshot times are given below of each figure.}
	\label{fig:fig8}
\end{figure}
We look at snapshots of radial magnetic field at outer boundary of the
spherical shell of different time instants of the nonlinear simulations.
Fig. \ref{fig:fig7} shows total Elsasser number of dynamo models at
two different $E$. 
Fig. \ref{fig:fig8} shows snapshots of radial magnetic field 
at outer boundary for nonlinear dynamo simulations. 
Very begining of the simulations both show a dipolar strcuture at the outer 
boundary. 
As time progess both dynamos show a multipolar structure. 
We further integrate our simulations to see if any difference arises. 
We observe an interesting and quite distinct behavior in the nonlinear simulations.
After some time the magnetic field switch back to dipolar strcuture at the outer 
boundary of spherical shell and it happens much earlier than the saturation stage. 
Though one would not expect this kind of behaviour in kinematic models. 
The magnetic field shows a first dipolar structure at 
$t_d=0.522$ and $t_d=0.329$ at $E=1.2 \times 10^{-5}$
and $E=1.2 \times 10^{-6}$, respectively (see, Fig. \ref{fig:fig7}, 
point \textquoteleft $d$ \textquoteright and \textquoteleft $i$ \textquoteright).
The corresponding total Elasser numbers are $0.42$ and $0.59$,  
which is well before saturation of the raidly rotating nonlinear dynamos. 
Therefore, this unique behavior of the nonlinear dynamo is solely 
due to the back reaction of the magnetic field. 
This event can be better visualise in a video. 
This behavior suggests that back reaction able to modify structure 
of the flow (therefore, helicity associates with the columnar flows) 
and bring back the dipolar srtucture well before saturation of the dynamo. 
This study suggests that the magnetic field can effect the flow much earlier 
than the saturation, even though Elsasser number is not high enough.
\par
\begin{figure}[t!]
	\centering
	\includegraphics[width=0.90\textwidth]{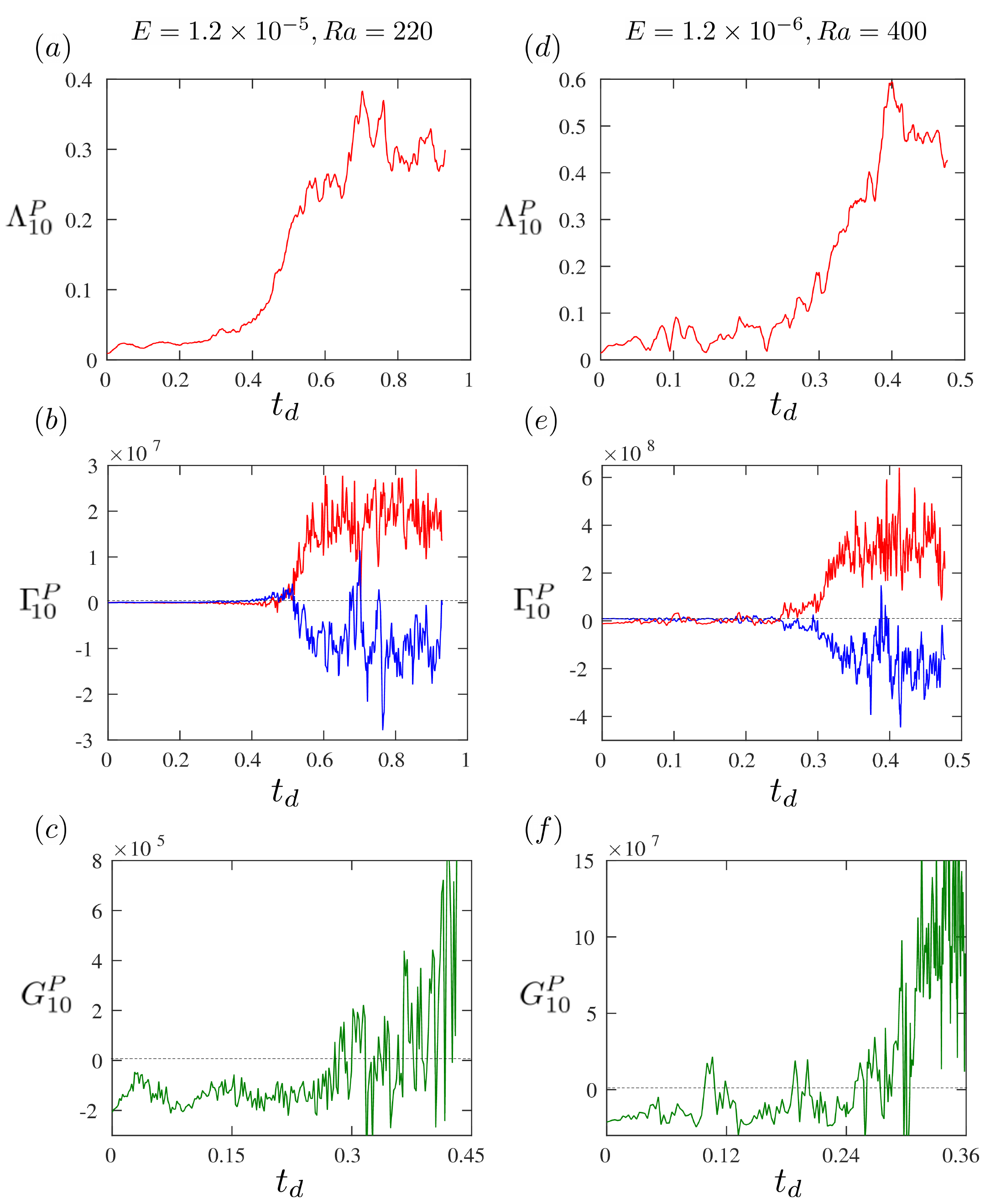}
	\caption{$(a)$ and $(d)$  show evolution of Elsasser number of $\bm{B}_{10}^P$ 
	for dynamo simulations at two different E. 
	$(b)$ and $(e)$ show energy contribution to $\bm{B}_{10}^P$. 
	Red line represent 
	$ (Pm/E) \bm{B}_{10}^P$.$\nabla \times (\bm{u}^T \times \bm{B}^P)$;
	Blue line represent 
	$ (Pm/E) \bm{B}_{10}^P$.$\nabla \times (\bm{u}^P \times \bm{B}^T)$ .
	($c$) and ($f$) show
	difference of red and blue lines.  }
	\label{fig:fig9}
\end{figure}
Fig. \ref{fig:fig9} shows evolution of Elsasser number based on 
axial dipole ($\Lambda^P_{10}$) and generation of axial dipole at 
two different $E$. 
Note that te initial Elsasser number is very small
($\Lambda^P_{10}=0.01$). 
The motivation is to trace the effect of the back reaction through the 
axial dipole generaion. 
The middle panel of Fig. \ref{fig:fig9} shows the contribution to the 
magnetc energy of $B^P_{10}$ by flow-field interaction in the induction equation. 
Indcution term is separated into four combinatons of toroidal-poloidal 
decompositions of flow and field.
Out of these four we see only two of them 
($\nabla \times (\bm{u}^T \times \bm{B}^P)$ and 
$\nabla \times (\bm{u}^P \times \bm{B}^T)$)  
are contributing in the generation. 
The red line represents energy coming from 
$\nabla \times (\bm{u}^T \times \bm{B}^P)$ and 
blue line shows energy coming from $\nabla \times (\bm{u}^P \times \bm{B}^T)$. 
The two plots show that initially blue line is positive and red
is negative and after a while they changed signed. 
This bifurcation point indicates when effect of back reaction becomes important 
in each dynamo simualation. 
Hence, we see a clear transition between kinematic 
and nonlinear phase in a self-consistent nonlinear dynamo model. 
Though before the bifuraction (kinematic phase) the values are too small
to see them in magnitude wise. 
To see which term dominates in this phase, we substract blue line from red line 
and the difference is defined as $G^P_{10}$. 
\begin{equation}\label{eq11}
\begin{aligned}
G^P_{10} = \int_{V} \biggl( \bm{B}^P_{10} \cdot \nabla \times (\bm{u}^T \times \bm{B}^P) -
\bm{B}^P_{10} \cdot \nabla \times (\bm{u}^P \times \bm{B}^T) \biggr ) dV
\end{aligned}
\end{equation}
The bottom panel of Fig. \ref{fig:fig9} shows zoomed of the $G^P_{10}$ for the 
two dynamo models. It suggests that initially $G^P_{10}$ is negative and then 
becomes positive. This ngetaive value shows the dominance of 
$\nabla \times (\bm{u}^P \times \bm{B}^T)$ in the kinematic phase. 
This refers to a transitional behaviour in the solution in the presence of the 
Lorentz force.
By visualization we can see the difference of higher magnitude 
of red line than blue line after the bifurcation till saturation of the dynamo. 
This confirms a positive growth of the axial dipole after the bifurcation. 
In the saturation state, the net postive value will balance by negative value
of the dissipation of the axial dipole. 
\par 
\begin{figure}[t!]
	\centering
	\includegraphics[width=0.85\textwidth]{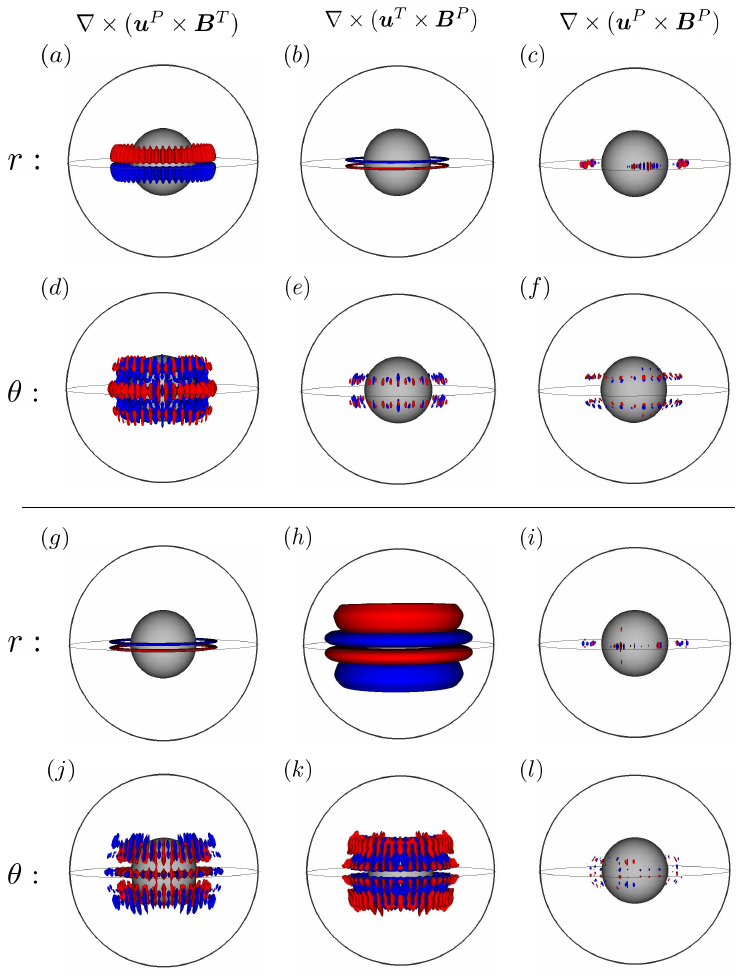}
	\caption{Isosurfaces of $r$ and $\theta$ components of induction terms for 
	nonlinear simulation at $E=1.2 \times 10^{-6}$, $Pr=Pm=1$, $Ra=400$. 
	Two snapshots are separated by horizontal line. 
	The top $(a-f)$ is at $t_d=0.2$  and bottom $(g-l)$ is at $t_d=0.38$.  
    The contour levels for $(a-f)$ are $\pm 5$ and $(g-l)$ are $\pm 20$. }
	\label{fig:fig10}
\end{figure}
The dynamo model consists of full spectrum of the flow and magnetic field. 
Therefore, it is difficult to visualize how small scale flows and fields are 
interacting with each others and producing a large scale magnetic field 
(herein, axial dipole) as suggested by mean field theory \cite{Steenback1966}. 
In this case, Bullard's selection works quite good way to help in 
finding the scale of flow and field and to see their interaction. 
The initial stage of the nonlinear simulation behaves as kinematic. 
Therefore, we compare two snapshot times, one in early phase and another is
further in time (before saturation of the dynamo)  of the simulation
to see the differences in mechanisms to generate the axial dipole in the simulation. 
For that, we choose two snapshots at $t_d=0.2$ (kinematic phase) and $t_d=0.55$ 
(nonlinear phase). 
Fig. \ref{fig:fig10} shows isosurafce snapshots of  
$\nabla \times (\bm{u}^P \times \bm{B}^T)$, 
$\nabla \times (\bm{u}^T \times \bm{B}^P)$ 
and $\nabla \times (\bm{u}^P \times \bm{B}^P)$
for nonlinear simulation
at $E=1.2 \times 10^{-6}, Ra=400$ at twp different snapshot times
($t_d=0.2$ and $0.38$) separated by a horizontal line. 
The values of $l$ and $m$ are truncated at $30$ while using Bullard's 
selection rule.
Below $l$ and $m=30$, there is no signifacent differences in structures. 
Fig. \ref{fig:fig10} ($a,b$ and $e,f$) show the $r$ components and
($c,d$ and $g,h$) show the $\theta$ components of the decompositions.
Comparison across $r$ and $\theta$ components of induction terms
above the horizontal line suggests that it is induction term 
$\nabla \times (\bm{u}^P \times \bm{B}^T)$ that is contributing 
to the axial dipole generation. 
The interaction of small scale flow and field is clearly evident 
in the dominant induction term.
For $r$ component it is antisymmeric and for $\theta$ component 
it is symmetric, though there is a sign change from negative to
positive across radial direction which also present in the
$\theta$ component of the axial dipole.
Figures below the line show it is $\nabla \times (\bm{u}^T \times \bm{B}^P)$ 
that is positively contributing to the both $r$ and $\theta$ 
components of the axial dipole. 
Though in the $\theta$ component it is $\nabla \times (\bm{u}^T \times \bm{B}^P)$ 
which is positively correlate with axial dipole while
$\nabla \times (\bm{u}^P \times \bm{B}^T)$ is negatively
correlate. 
Therefore in the energy analysis, we see a postive contribution from 
$\nabla \times (\bm{u}^T \times \bm{B}^P)$ and negative from 
$\nabla \times (\bm{u}^P \times \bm{B}^T)$ as time progress. 
However the $r$ component of $\nabla \times (\bm{u}^T \times \bm{B}^P)$ 
shows a large scale axisymmetric structure and having a sign of change
across each hemisphere. 
Since we use Bullard's selection to rule
to calculate the terms for axial dipole, so there is a overlap of range
of $l$ which contribute to other harmonics of magnetic fields (for example,
octupole).   
Though in both snapshot times $\nabla \times (\bm{u}^P \times \bm{B}^P)$ is 
negligible.
\par
\begin{figure}[t!]
	\centering
	\includegraphics[width=\textwidth]{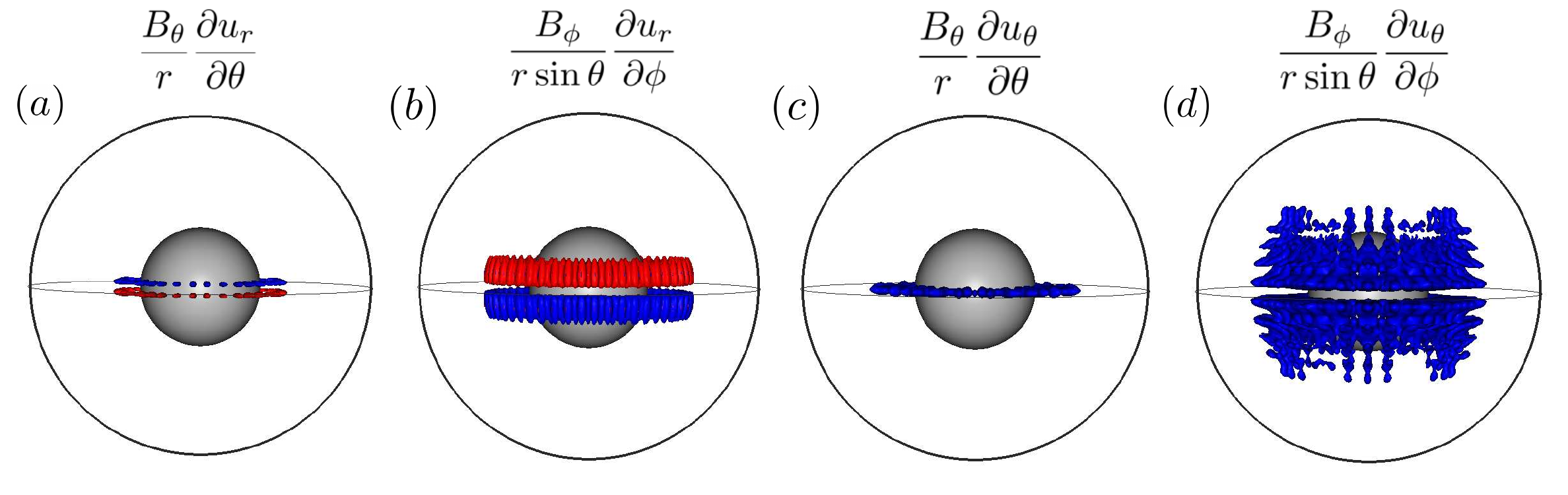}
	\caption{Isosurfaces of decompositions of $ \nabla \times (\bm{u}^P \times \bm{B}^T)$  	
	    for nonlinear simulation at $E=1.2 \times 10^{-6}$, $Pr=Pm=1$, $Ra=400$. 
	    The snapshot time is at $t_d=0.2$. The contour levels is $\pm 5$. }
	\label{fig:fig11}
\end{figure}

Now we look into components of the induction term through axial dipole
is getting generated. 
So for this we expand the induction in terms of stretching and advection.
The expanded form of $\nabla \times (\bm{u}^P \times \bm{B}^T)$ is
\begin{equation}\label{eq12}
\begin{aligned}
\nabla \times (\bm{u}^P \times \bm{B}^T) = 
\biggl[\underbrace {\frac{B_\theta}{r} \frac{\partial u_r}{\partial \theta}}_\text{I} 
+ \underbrace{\frac{B_\phi}{r \sin \theta} \frac{\partial u_r}{\partial \phi}}_\text{II}
 \biggr] \widehat{\bm{e}}_{r} \\
+ \biggl[ \underbrace {\frac{B_\theta}{r} \frac{\partial u_\theta}{\partial \theta}}_\text{III}  
+ \underbrace{\frac{B_\phi}{r \sin \theta} \frac{\partial u_\theta}{\partial \phi}}_\text{IV} 
+ {\frac{B_\theta u_r}{r}} \biggr] \widehat{\bm{e}}_{\theta}  \\
- \biggl[ {u_r \frac{\partial B_\theta}{\partial r}}   
+{\frac{u_\theta}{r} \frac{\partial B_\theta}{\partial \theta}} 
+{\frac{u_\phi}{r \sin \theta} \frac{\partial B_\theta}{\partial \phi}} 
 \biggr]  \widehat{\bm{e}}_{\theta}
\end{aligned}
\end{equation}
Similarly, $\nabla \times (\bm{u}^T \times \bm{B}^P)$ is expand as
\begin{equation}\label{eq13}
\begin{aligned}
\nabla \times (\bm{u}^T \times \bm{B}^P) = 
- \biggl[ \underbrace{\frac{u_\theta}{r} \frac{\partial B_r}{\partial \theta}}_\text{V} 
+\underbrace{\frac{u_\phi}{r \sin \theta} \frac{\partial B_r}{\partial \phi}}_\text{VI} \biggr]
\widehat{\bm{e}}_{r} \\ 
+\biggl[ \underbrace{B_r \frac{\partial u_\theta}{\partial r}}_\text{VII}  
+\underbrace{\frac{B_\theta}{r} \frac{\partial u_\theta}{\partial \theta}}_\text{VIII} 
+{\frac{B_\phi}{r \sin \theta} \frac{\partial u_\theta}{\partial \phi}} \biggr]
\widehat{\bm{e}}_\theta  \\
-\biggl[ {\frac{u_\theta}{r} \frac{\partial B_\theta}{\partial \theta}}    
+{\frac{u_\phi}{r \sin \theta} \frac{\partial B_\theta}{\partial \phi}}
+{\frac{u_\theta B_r}{r}} \biggr] \widehat{\bm{e}}_\theta 
\end{aligned}
\end{equation} 
Table \ref{tab:decomp1} and Table \ref{tab:decomp2} show the volume averaged 
of over top hemisphere of induction terms ($\nabla \times (\bm{u}^P \times \bm{B}^T)$
and $\nabla \times (\bm{u}^T \times \bm{B}^P)$) and its decomposition as shown in 
Eq. \ref{eq12} and Eq. \ref{eq13} for two different time instants at
$t_d=0.2$ and $0.35$.
Fig. \ref{fig:fig11} shows the dominant terms of decompositions of 
$\nabla \times (\bm{u}^P \times \bm{B}^T)$ as defined in 
Eq. \ref{eq12}. The terms are calculated by Bullard selection rule. 
The snapshots are shown at $t_d=0.2$.
The $r$ and $\theta$ components of axial dipole are getting generated from 
deformation of the small scale $B_\phi$ by $u_r$ and $u_\theta$, respectively.
This suggests that for primary flow (rotation of the columns about its 
vertical axis) is responsible for $B_r$ and the seconadry flow (up and down
flow inside the column) is responsible for the generation of $B_\theta$
of the axial dipole \cite{Jones2007}. 
This argument is based on low-$Rm$
approximation, where first order smoothing approximation is relevant. 
\cite{Moffatt1995}.
Here length scale of $Rm$ is based on thickness of the roll.   
Nevertheless, we find that till $t_d=0.3$ the growth of axial dipole
is very weak for $E=1.2 \times 10^{-6}, Ra=400$ 
(see, Fig. \ref{fig:fig9} ($b$)).
The dominant structures of the stretching terms are similar to the dominant 
induction term as shown in Fig. \ref{fig:fig10}.
For radial field we can see the antisymmetric structure above and below of 
the equator plane, mimicing $Y^0_1$ strcuture of the poloidal field while
for $\theta$ component it is symmetric about the equator.
In both case the interaction between small scale structures flows and 
fields are notable.
\begin{table}[H]
\caption{Volume averaged over top hemisphere of r-component of induction 
        term and its decomposition as shown in Eq. \ref{eq12} and Eq. \ref{eq13}. 
        The values reported are here at two different time instants 
        ($t_d=0.2$ and $0.38$).}
    \begin{minipage}{0.5\textwidth}
      \centering
      \begin{tabular}{c c c}
        \toprule ${t_d}$ & $0.2$ & $0.38$   \\
        \midrule ${\nabla \times (\bm{u}^P \times \bm{B}^T)}_r$ & 10.36 & -0.98 \\
        $({B_\theta}/{r}) ({\partial u_r}/{\partial \theta}) $ & 8.36 & -0.54 \\
        $({B_\phi}/{r sin\theta})({\partial u_r}/{\partial \phi})$ & 2.05 & -0.44 \\
        \bottomrule
      \end{tabular}
    \end{minipage}
    \begin{minipage}{0.5\textwidth}
      \centering
      \begin{tabular}{c c c}
        \toprule ${t_d}$ & $0.2$ & $0.38$   \\
        \midrule  ${\nabla \times (\bm{u}^T \times \bm{B}^P)}_r$ & 1.36 & 24.98 \\ 
        $-({u_\theta}/{r}) ({\partial B_r}/{\partial \theta})$ & 0.56 & 23.74 \\
        $-({u_\phi}/{r sin\theta})  ({\partial B_r}{\partial \phi})$ & 0.75 & 0.54 \\
        \bottomrule
      \end{tabular}\label{tab:decomp1}
    \end{minipage}
  \end{table}
\begin{table}[H]
\caption{Volume averaged over top hemisphere of $\theta$-component of induction 
        term and its decomposition as shown in Eq. \ref{eq12} and Eq. \ref{eq13}. 
        The values reported are here at two different time instants 
        ($t_d=0.2$ and $0.38$)}
    \begin{minipage}{0.5\textwidth}
      \centering
      \begin{tabular}{c c c}
        \toprule ${t_d}$ & $0.2$ & $0.38$   \\
        \midrule ${\nabla \times (\bm{u}^P \times \bm{B}^T)}_\theta$ & -12.36 & 20.54 \\
        $({B_\theta}/{r}) ({\partial u_\theta}/{\partial \theta})$ & -2.42 & 18.54 \\
        $({B_\phi}/{r sin\theta})({\partial u_\theta}/{\partial \phi})$ & -10.36 & 0.53 \\
        $-{u_r} ({\partial B_\theta}/{\partial r}) $ & 0.36 & 0.42 \\
        $-({u_\theta}/{r})({\partial B_\theta}/{\partial \theta})$ & 0.22 & 0.18 \\
        $-({u_\phi}/{r}) ({\partial B_\theta}/{\partial \phi}) $ & 0.18 & 0.22 \\        
        \bottomrule
      \end{tabular}
    \end{minipage}
    \begin{minipage}{0.5\textwidth}
      \centering
      \begin{tabular}{c c c}
        \toprule ${t_d}$ & $0.2$ & $0.38$   \\
        \midrule ${\nabla \times (\bm{u}^T \times \bm{B}^P)}_\theta$ & 0.22 & -30.54 \\ 
        ${B_r} ({\partial u_r}/{\partial \theta})$ & 0.12 & -24.62 \\
        $({B_\theta}/{r})({\partial u_\theta}/{\partial \theta})$ & 0.16 & -2.94 \\
        $({B_\phi}/{r sin\theta}) ({\partial u_\theta}/{\partial \phi}) $ & 0.21 & -1.57 \\
        $-({u_\theta}/{r})({\partial B_\theta}/{\partial \theta})$ & 0.13 & 0.94 \\
        $-({u_\phi}/{r}) ({\partial B_\theta}/{\partial \phi}) $ & 0.17 & 0.83 \\ 
        \bottomrule
      \end{tabular}\label{tab:decomp2}
    \end{minipage}
\end{table}
\begin{figure}[t!]
	\centering
	\includegraphics[width=\textwidth]{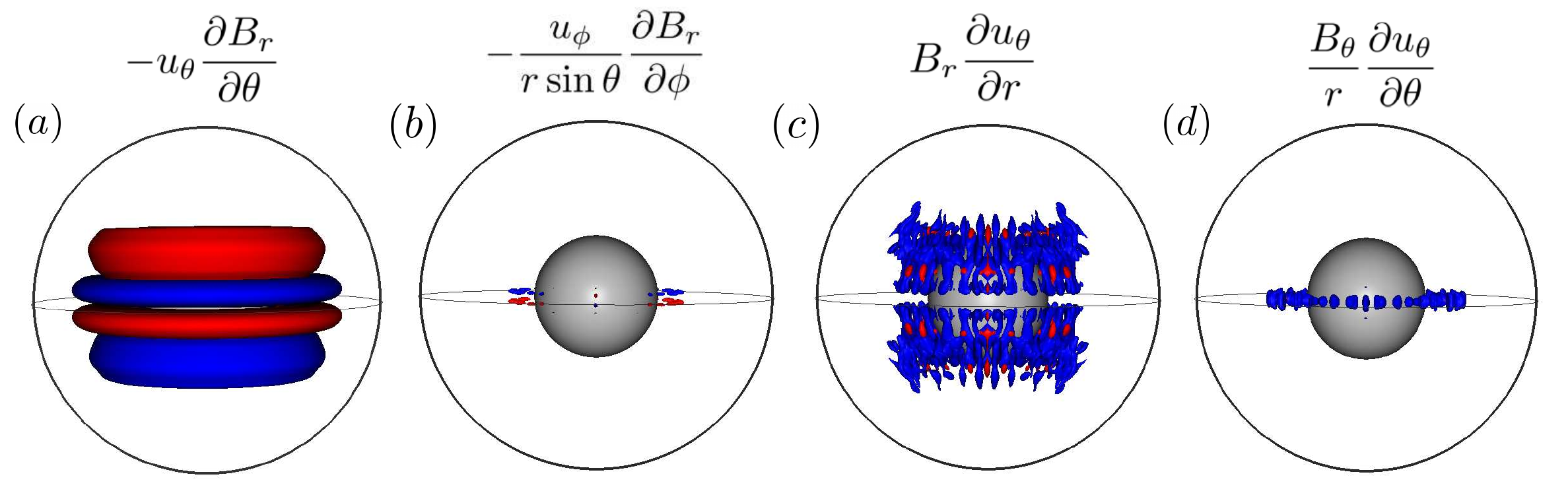}
	\caption{Isosurfaces of decompositions of 
	$ \nabla \times (\bm{u}^T \times \bm{B}^P)$	
	    for nonlinear simulation at $E=1.2 \times 10^{-6}$, $Pr=Pm=1$, $Ra=400$. 
	    The snapshot time is at $t_d=0.38$. The contour levels is $\pm 20$. }
	\label{fig:fig12}
\end{figure}
\par
Though the above mechanism is not valid as the magnetic field effects the flow.
Fig. \ref{fig:fig12} shows the dominant terms of decompositions of 
$\nabla \times (\bm{u}^T \times \bm{B}^P)$ as defined in 
Eq. \ref{eq13}. The terms are calculated by Bullard selection rule.
The snapshots are shown at $t_d=0.38$.
Fig. ($a, b$) and ($c, d$) show $r$ and $\theta$ component of
the induction term.
The dominant contributions for $r$ and $\theta$ components 
of the axial dipole is coming from 
$-(u_\theta/r) {\partial B_r}/{\partial \theta}$  and 
$B_r {\partial u_\theta}/{\partial r}$. 
Note that the dominant structures of the terms are similar to the  
induction terms as shown in Fig. \ref{fig:fig10}
The $r$ component is supported by a advection of the radial field.
Though, advection does not contribute in overall magnetic energy. 
The term analysis of $\nabla \times (\bm{u}^T \times \bm{B}^P)$  
(refer to Eq. \ref{eq13}) suggests that there is no stretching term to
contribute to the radial component of axial dipole. 
Hence, this is a unique behaviour of nonlinear simulations. 
Though we have not looked into the transport coefficeints of mean field 
theory using test field method as done by many authors in rotating 
spherical shell dynamos 
\cite{Schrinner2005} \cite{Schrinner2006} \cite{Schrinner2007}. 
The advection of mean magnetic flux by a mean flow is denoted by $\bm{\gamma}$ 
(see Eq. 11  of \cite{Schrinner2007}). 
Schrinner et. al \cite{Schrinner2007} showed that out of three combination 
of $\bm{\gamma}$, $\gamma_{\theta}$ is contributing more in the
generation of mean field. 
$\gamma_{\theta}$ represents advection of the mean field in $\theta$ direction. 
The coefficients of mean field suggests that the advection played 
an important role in preference of dipole in the context of rotating 
nonlinear dynamos \cite{Schrinner2007}. 
Nevertheless, this strcuture of advection is similar to the advection of 
the mean field in meriodinal direction ($\gamma_\theta$) as shown by 
Schrinner et. al \cite{Schrinner2012}. 
Fig. \ref{fig:fig11}  shows the energy contribution to $\theta$ 
component of the axial dipole. 
It shows that the dominant contribution is coming from the stretching term 
$B_r {\partial u_\theta}/{\partial r}$. 
The $\theta$ component of the axial dipole is contributed by 
$B_r {\partial u_\theta}/{\partial r}$. 
Note that this term is zero at equator because $B_r$ is zero.
Though meriodinal component of axial dipole is not zero and therefore, 
we see a non negligible contribution from 
$(B_\theta/r) {\partial u_\theta}/{\partial r}$.


\subsubsection{Effect of Lorentz force in energy transfer}

\begin{figure}[t!]
	\centering
	\includegraphics[width=\textwidth]{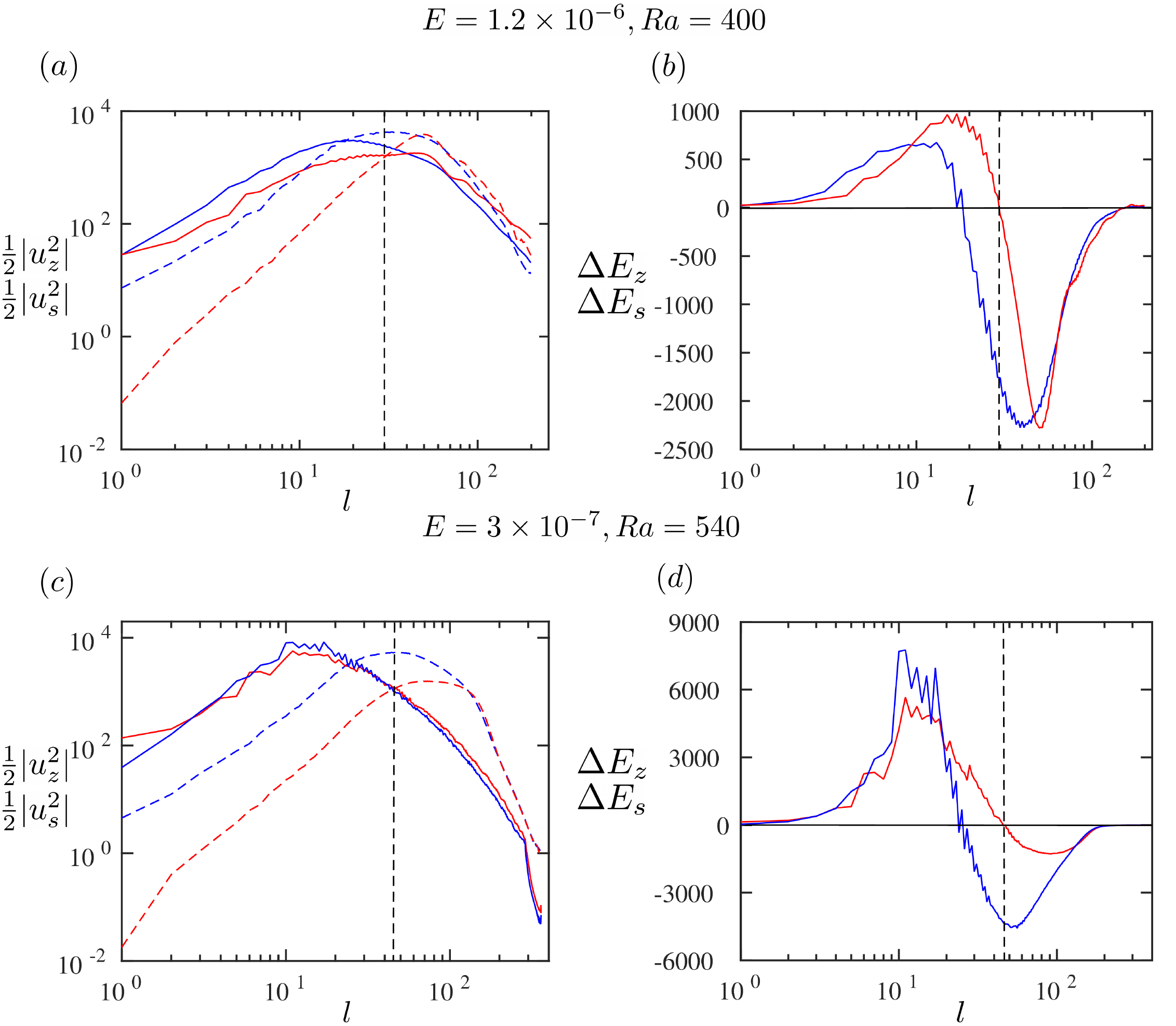}
	\caption{$(a)$ and $(c)$ show time averaged of $z$-kinetic energy (red) and 
	$s$-kinetic energy (blue) spectra over harmonic degree $l$ for nonlinear and 
	non-magnetic simulations at two different $E$. 
	Solid line is for nonlinear dynamo simulation and dotted line is for non-magnetic 
	simulation. 
	The vertical dotted line shows where $z$-kinetic energy of nonlinear simulation 
    falling below nonmagnetic simulation.
	$(b)$ and $(s)$ show the differences in magnitude of $z$-kinetic energy (solid red 
	line) and $s$-kinetic energy (solid blue line) between nonlinear and non-magnetic 
	simulation over $l$. 
	For $E=1.2 \times 10^{-6}$, the vertical line is at $l=30$
	and for $E=3 \times 10^{-7}$, the vertical line is at $l=46$. }
	\label{fig:fig17}
\end{figure}
Next, we look into effect of the magnetic field on the flow. One of the way
to quantify to look into energy spectra in the satuarted phase
and compare between nonmagnetic and nonlinear dynamo models. 
To see effect of the magnetic field, time averaged of $u_z$ and $u_s$ is 
calculated in both saturated phase of non-magnetic and nonlinear models.
Fig. \ref{fig:fig17} ($a$  and $c$) shows time averaged of $u_z$ and $u_s$ at 
two different $E$. 
The dotted and solid lines (red - $u_z$ and blue - $u_s$) shows saturated 
non-magnetic and nonlinear models, respectively.
Fig. ($b$  and $d$) shows difference of kinetic energy between
nonlinear and non-magnetic models as defined as $\Delta E$. 
Comparison between nonlinear dynamo and non-magnetic energy spectra suggests
that the Lorentz force acts different way in different scales. 
Enhancement of $u_z$ and $u_s$ (for example, $l$ $leq$ $30$ at $E=1.2 \times 10^{-6}$) 
in case of nonlinear dynamo over non-magnetic case suggest that the Lorentz force 
acts to relax the geostrophic constraint imposed by the rotation, prefering a scale 
of convection larger than in absence of the magnetic field \cite{Soward1979}. 
Though, in the small scale, transfer of the kinetic to magnetic energy
through Lorentz force causes a lower kinetic energy than non-magnetic.
Traditional idea is that the effect of Lorentz force is only visible
when the nonlinear dynamo model satuartes. 
However, our results suggests that Lorentz effect the flow much before 
the saturation. 
Though how the dynamo models enter into nonlinear equillibrium state is 
not well understood at present.
This requires further regressive study to answer the fundamental questions of
turbulent dynamo models. 
There are few mathematical models, like $\alpha$-queching
mechanism \cite{Rudiger1993} or suppresion of Lagrangian chaos of the flow 
\cite{Cattaneo1996}, which shed some light on it. 
Though, there has been no specific mathematical models to answer it 
for a conevction driven turbulent dynamo models. 
It is known that for $\Lambda \sim \mathcal{O}(1)$, Magnetic-
-Archimedian-Coriolis (known as, MAC) forec balance holds in 
a saturated state \cite{Taylor_1963}. 
\par
\begin{figure}[t!]
	\centering
	\includegraphics[width=\textwidth]{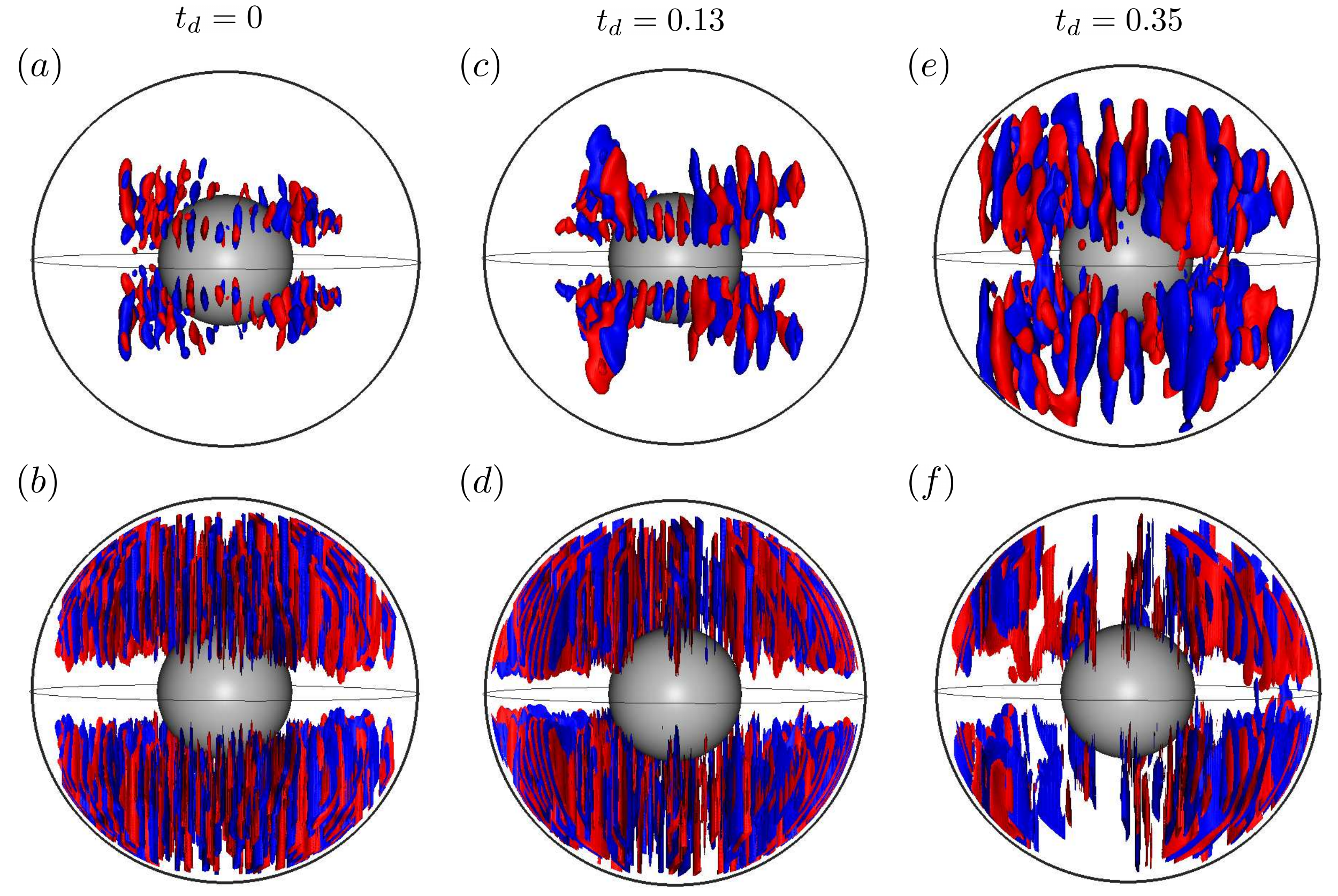}
	\caption{Top panel ($a, c$ and $e$) shows isosurface of $u_z$ (contour levels $\pm 80$) 
	   in the range of $l$ $\leq$ $30$ for nonlinear simulation at $E=1.2 \times 10^{-6}, 
	   Pr=Pm=1, Ra=400$. Bottom panel ($b, d$ and $f$)shows isosurface of $u_z$ 
	   (contour levels $\pm 150$) in the range of $l$ $>$ $30$ at the same parameters. }
	\label{fig:fig19}
\end{figure}

However in this study our aim is to show the effect of Lorentz
force much earlier than saturation. 
We look into $u_z$ at different time instant in the growing phase of 
the nonlinear dynamo model.  
Fig. \ref{fig:fig19} shows three snapshots of isosurafces $u_z$ in 
spherical shell at two spectra range (top panel - $l$ $\leq$ $30$ and 
bottom panel - $l$ $>$ $30$) for $E=1.2 \times 10^{-6}$. 
The last snaphots are taken during the growth phase of the
magnetic field. 
Top panel shows that as time progess there is generation of convection
outside the TC which is quiescent otherwise in the absence of the
magnetic field.
This implies that the Lorentz force enhances the
convection, which in turn further ampilifies the magnetic field
in the context of a self-sustained dynamo model.
This effect is a hallmark of large scale dipolar magnetic field, 
(in dynamo model it is axial dipole) which enhances the columnar 
flow \cite{Sreeni2011}.
However in the bottom panel shows decrease in amplitude of $u_z$
with time.
This effect resembles transfer of magnetic energy from kinetic energy
the field grows, which is a classical MHD results and the basis of 
dynamo action \cite{Davidson2001}. 
\par
\begin{figure}[t!]
	\centering
	\includegraphics[width=\textwidth]{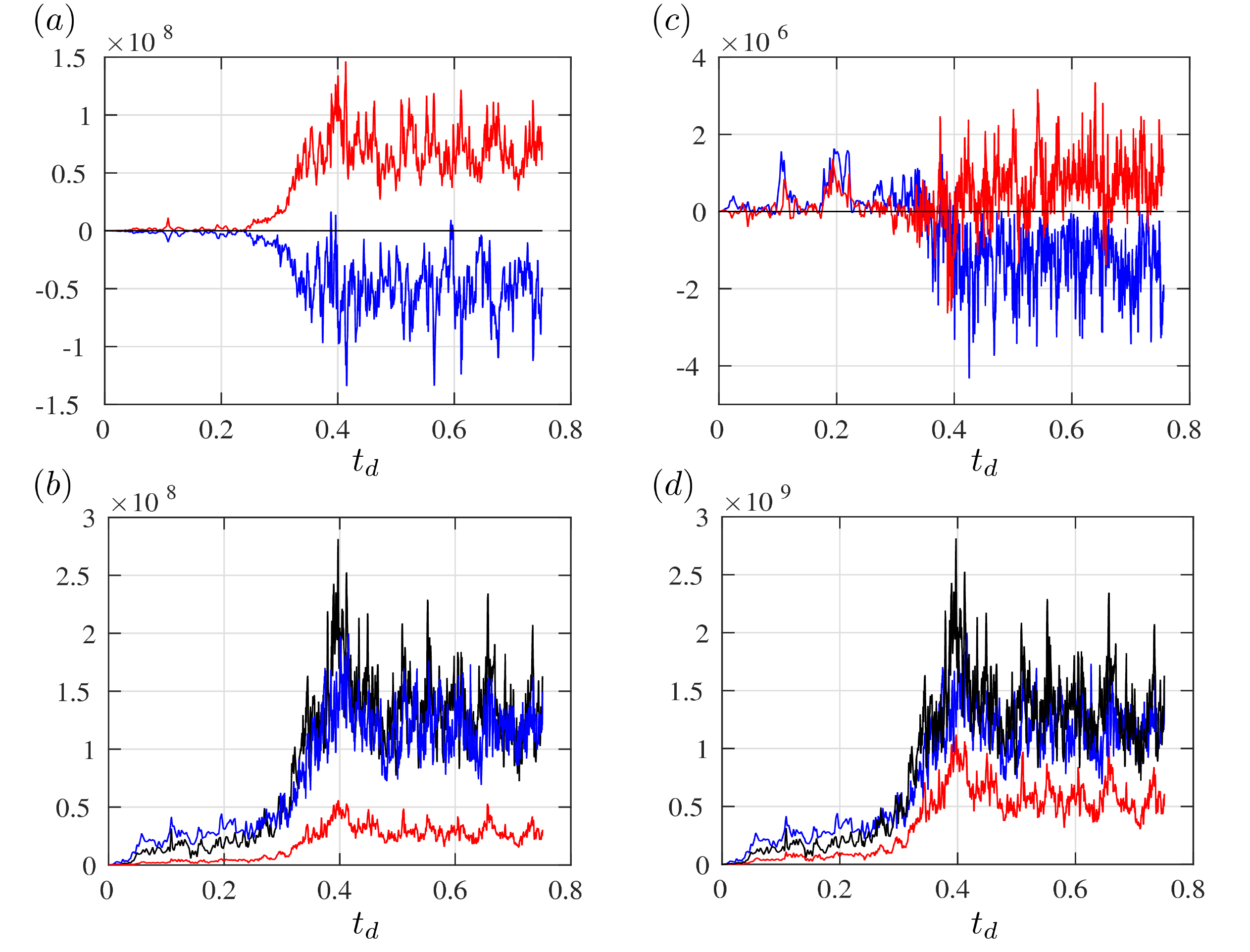}
	\caption{Work done by Lorentz force in the two ranges of $s$ and 
	$z$-kinetic energy spectra. ($a$ and $c$) represent work done by the Lorentz force based on axial     
    dipole in the nonlinear simulation. ($b$ and $d$) represent work done by the Lorentz force based on 
    nondipole components. For ($a$ and $b$) truncated value of $l$ for the flow is $30$. ($c$ and $d$) 
    shows $l$ from $30$ to $l_{max}=220$. Red line represents energy transfer by Lorentz 
        force from $u_s$, blue line represents from $u_z$ and black line represents 
    from $u_\phi$.  }
	\label{fig:fig18}
\end{figure}

Therefore we look into energy transfer by the Lorentz force in the $s$ and
$z$-momentum equations. 
Two types of Lorentz force consider here - Lorentz force based on axial dipole 
and nondipole. 
Lorentz force based on axial dipole is defined as 
$(\nabla \times \bm{B}^P_{10}) \times \bm{B}$ 
where, $\bm{B}^P_{10}$ is axial dipole and $\bm{B}$ consider the full spectrum.
Similar way, nondipole Lorentz force is defined where nondipole component
(except axial dipole) consider within the bracket. Energy transfer by Lorentz force
is defined as 
\begin{equation}\label{en_tra}
\begin{aligned}
\Sigma_j^i = -\int_{V} \bm{u}_i \cdot [(\nabla \times \bm{B}_j)  \times \bm{B}]  dV
\end{aligned}
\end{equation}
where $\Sigma_j^i$ represents volume average transfer over spherical shell of $i$-th 
component of velocity field to $j$-th component of magnetic field by Lorentz force 
\cite{Takahashi2012}.  
Energy transfers from two different ranges of energy spectra are shown in 
Fig. \ref{fig:fig18} based on two different Lorentz forces.
Fig. ($a$) and ($b$) shows energy transfer from velocity (red - $u_z$,
blue - $u_s$  and black - $u_\phi$) by axial dipole Lorentz force
for $l$ $\leq$ $30$ and $l$ $>$ $30$. 
Notable that energy transfer from $u_\phi$ is zero in both
ranges which means there is no role of $u_\phi$ in axial dipole 
generation. 
Fig. ($a$) suggests that work done by axial dipole Lorentz force in $u_z$ is
positive and for $u_s$ is negative. 
This implies that there Lorentz force is taking energy from $u_z$ and giving 
to $u_s$. 
Therefore this taking-giving role implies a self-propagation of the
axail dipole in a self-sustained dynamo model.  
However the important note here is that this happens much eariler the saturation,
when the magnetic field is weak.   
Though, the action of Lorentz force on the small scale is comparatively weak
as shown in Fig. ($b$). The mean values in the satuarted state
for $u_z$ in the range of $l$ $\leq$ $30$ and $l$ $>$ $30$ are
$6.93 \times 10^{7}$ and $6.95 \times 10^{5}$.  
The mean values in the satuarted state
for $u_s$ in the range of $l$ $\leq$ $30$ and $l$ $>$ $30$ are
$-5.41 \times 10^{7}$ and $-1.45 \times 10^{6}$.
\begin{figure}[t!]
	\centering
	\includegraphics[width=0.6\textwidth]{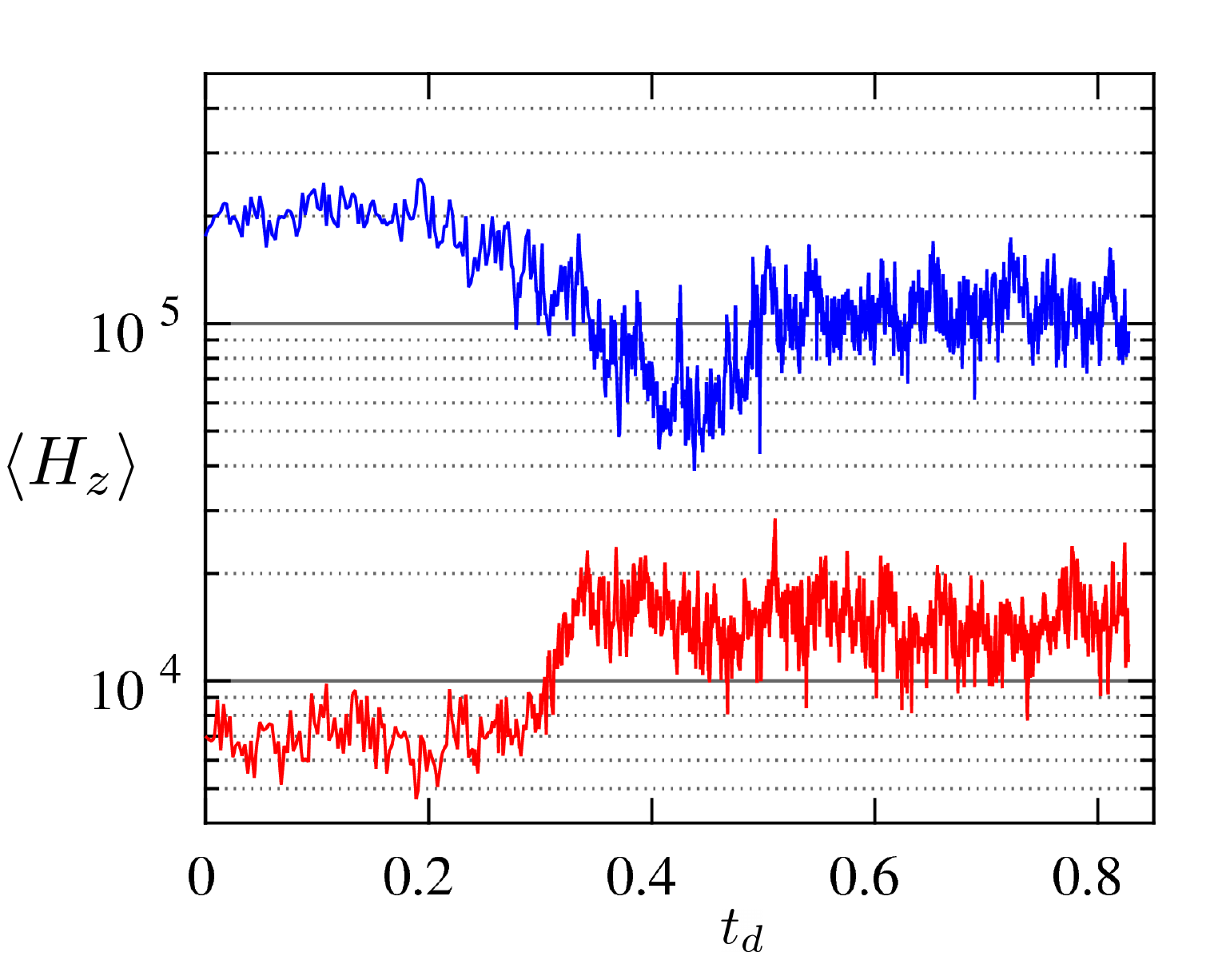}
	\caption{Evolution of volume averaged of $z$-helicity ($ \langle H_z \rangle$) over 
	       southern hemisphere with time. Red line represents $\langle H_z \rangle$ in the 
	       range of $l$ $\leq$ $30$ and blue line represents $\langle H_z \rangle$ in the 
	       range of $l$ $>$ $30$ for nonlinear simulation at $E=1.2 \times 10^{-6}, Pr=Pm=1, 
	       Ra=400$.}
	\label{fig:fig20}
\end{figure}
\par
\subsubsection{Helicity generation and preferences in axial dipole magnetic field}
\begin{figure}[t!]
	\centering
	\includegraphics[width=0.8\textwidth]{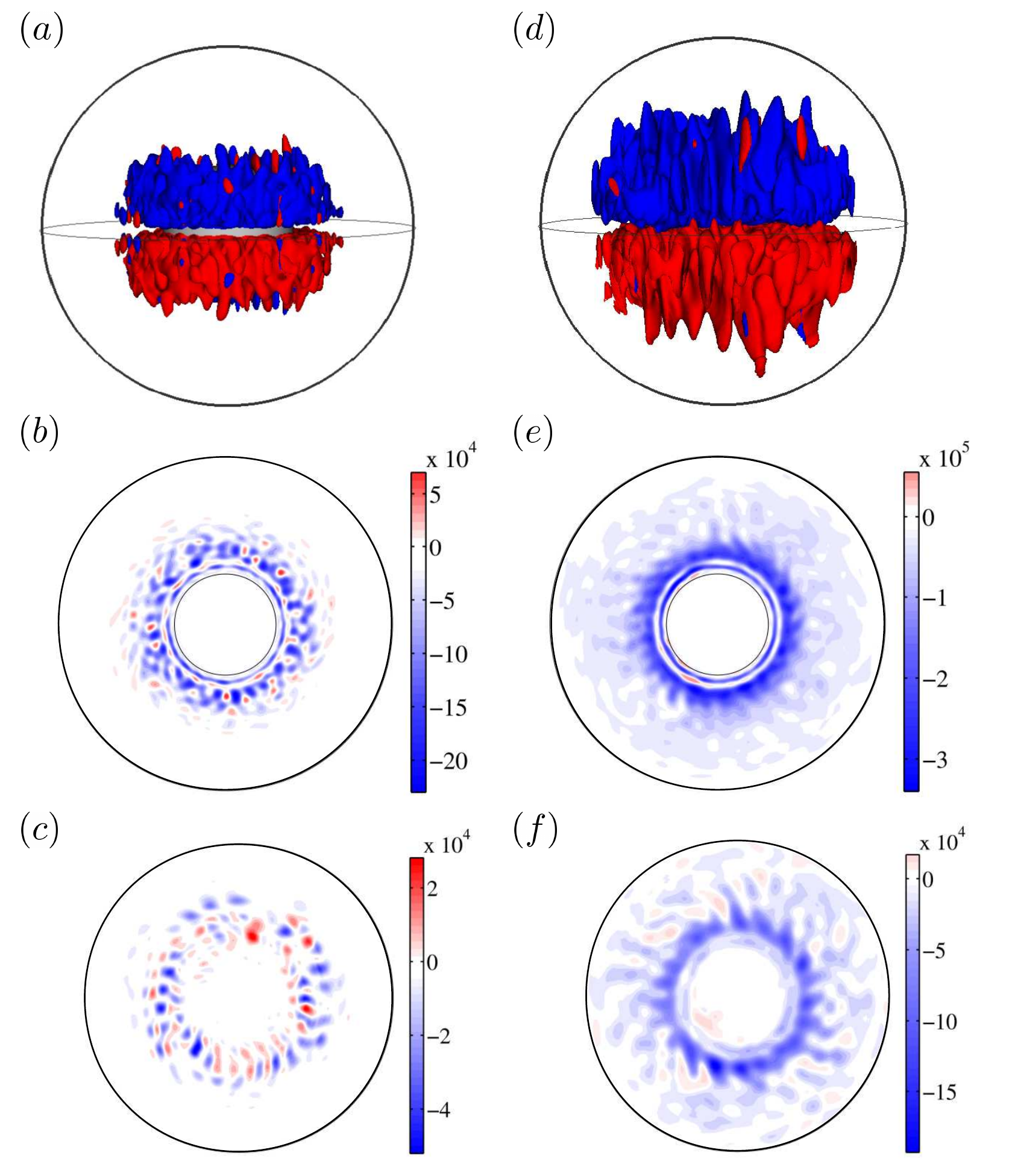}
	\caption{Top panel ($a$ and $d$) shows isosurface of $z$-helicity ($H_z$) 
	      (contour levels $\pm 2 \times 10^{4}$) in the range of 
		$l$ $\leq$ $30$ for simulations at $E=1.2 \times 10^{-6}, Pr=Pm=1, Ra=400$. 
		The left panel is for nonmagnetic and right panel is for nonlinear dynamo.
		Middle panel ($b$ and $e$) shows $H_z$ at $z$=0.3 and bottom panel ($c$ and $f$) 
		shows $H_z$ at $z$=0.6.   }
	\label{fig:fig21}
\end{figure}

The effect of Lorentz force on the flow in a nonlinear dynamo model
can be understand by looking at the helicity distribution as shown 
in rapidly rotating linear magnetoconvection in spehrical shell 
by previous study \cite{Sreeni2011}. 
The authors imposed a large scale azimuthal magnetic field and find an 
enhancement of $z$-helicity in case of imposed dipolar magnetic 
field over a quadrupolar field under rapidly rotation, although 
strength of the field is small.
This happens due to the fact that the Lorentz force enhance the 
columnar flow by breaking Taylor-Proudman constrain.
However in rapidly rotating dynamo models suggests that the length
scale of axisymmetric azimuthal field in the direction of rotation
decreases by lowering $E$ \cite{Venk2015}.  
The authors has shown that by lowering the length scale of the
magnetic field, the enhancement of the helicity is getting reduced 
compare to the non-magnetic case.
Therefore care must be taken to make a direct comparison of magnetoconvection 
studies with nonlinear models.
The essenatial difference that comes from magnetoconvection
studies that the results are dependent on the initial structure
of the imposed field.
However the structure of magnetic field in nonlinear models are
see too complicated to scale dependent effect of the field on the 
flow.
For that we look into helicity in two different scale of the flow.
Fig. \ref{fig:fig20} shows evolution of the volume averaged $z$-helicity
(after removing the boundary layer at two boundaries) over northern
hemipshere at two different length scales for nonlinear dynamo model
at $E=1.2 \times 10^{-6}$. 
Red line shows $z$-helicity of the flow $l \leq 30$ and blue shows for $l < 30$.
Overall enhancement of the $z$-kinetic helicity with time as magnetic field grows is one of
most striking behaviour of the large scale magnetic field as linear theory predicts.
In small large scale we see that $H_z$ associated with the flow decreases
as field grows. 
The initial and mean value at saturated $z$-helicity for $l \leq 30$ are 
$7 \times 10^3$ and $1.39 \times 10^4$, respectively. 
There is almost $2$ times of enhancement due to large scale magnetic field. 
This happens due to the fact that there is a transfer of kinetic
energy to magnetic field by the Lorentz force.
Fig. \ref{fig:fig21} shows comparison of time averaged of $H_z$ between non-magnetic 
and magnetic simulations for $l \leq 30$. 
Fig. \ref{fig:fig21} ($a$) and ($d$) show isosurafce of $H_z$ in spherical shell
at contour level of $\pm 2 \times 10^4$ for non-magnetic and magnetic
case, respectively. 
For non-magnetic case, $H_z$ is confined mostly to inner boundary, while in magnetic 
case $H_z$ spread over to the outer region. 
Hence there is an overall enhancement of $H_z$ as in the range of
$l \leq 30$ as we see in Fig. \ref{fig:fig20}.
The middle and bottom panels shows the same plot at two different $z$-sections
(at $z=0.3$ and $0.6$).     
In the $z$-section plot, we see that there is a both postive and negative
helicity contribution in the non-magnetic case and while, in magnetic there
is a coherent structure of negative helicity present.
Fig. \ref{fig:fig21} shows cyclone and anticyclone for two ranges of $l$, which are $l \leq 30$ and $l > 30$. 
Interstingly, the helicity asymetry shows up only at $l > 30$ for the nonlinear 
dynamo models, while non-magnetic convective models does not shows up any such behaviour.
However, the range of $l \leq 30$, there is no such asymetry between cyclone and anticyclone 
helicity in nonlinear dynamo model, although a overall enhancement is seen compare to non-magnetic 
case. This behaviour clearly shows the scale dependent behaviour of the magnetic field on the
helicity, hence on the flow. 
\begin{figure}[t!]
	\centering
	\includegraphics[width=\textwidth]{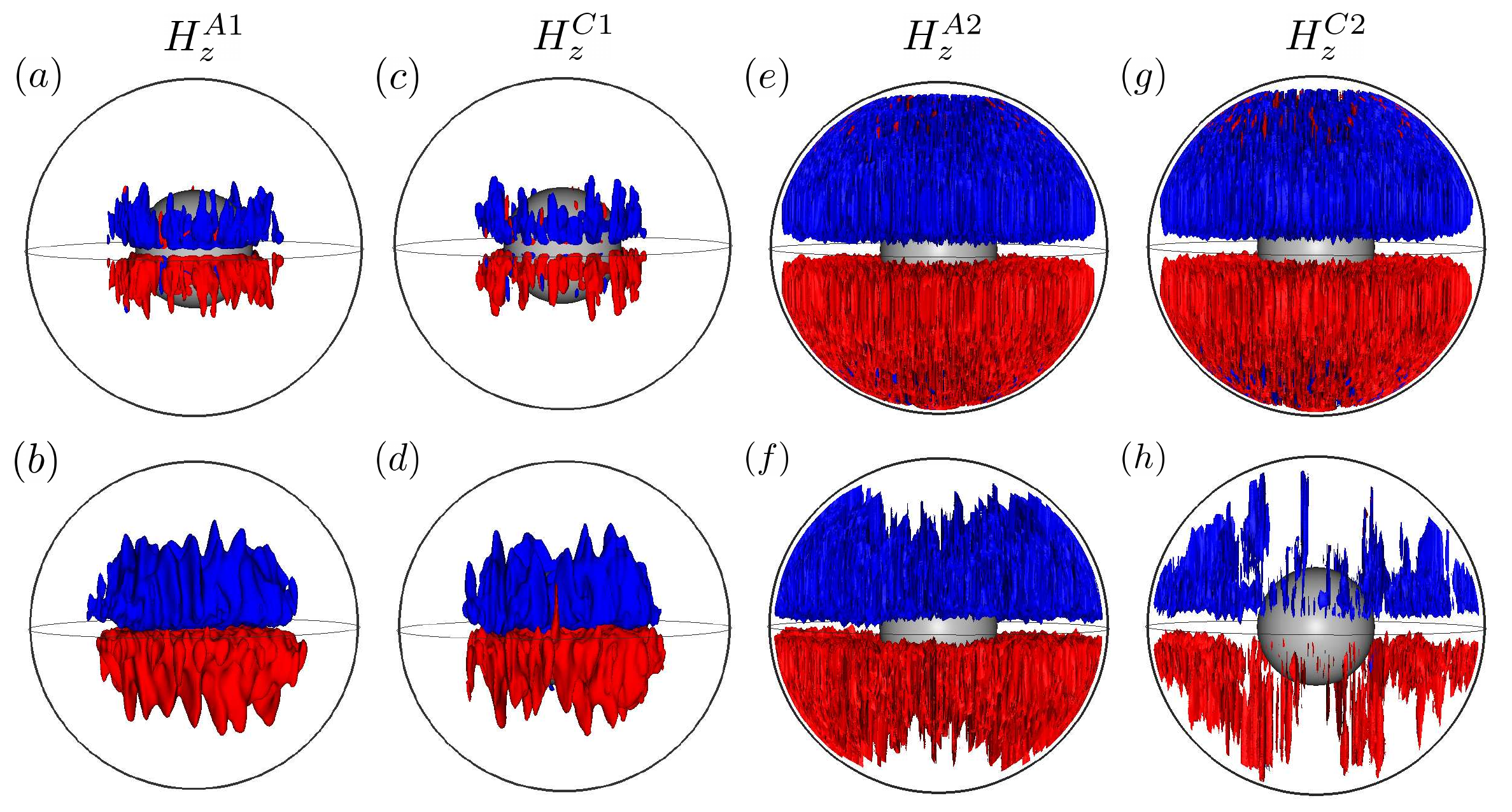} 
	\caption{Top panel shows isosurfaces of time averaged of anticyclonic (A) and 
	    cyclonic (C) $z$-helicity for nonmagnetic simulation and bottom panel shows 
	    for nonlinear dynamo simulation at $E=1.2 \times 10^{-6}, Pr=Pm=1, Ra=400$. 
	    ($a - d$) represent in the range of $l$ $\leq$ $30$ (contour levels 
		$\pm 2 \times 10^{4}$ ) and ($e - h$) represent in the range of 
		$l$ $>$ $30$ (contour levels $\pm 2 \times 10^{5}$ ).  }
	\label{fig:fig22}
\end{figure}

\clearpage
\section{Discussion and conclusions}
In this study we investigate preferences of the axial dipole structure from a 
seed magnetic field in the rapidly rotating spherical shell dynamo models. 
We study both nonlinear as well as kinematic regimes. For kinematic models 
Lorentz force is droped from the momentum equation.  In rapidly rotating system, 
onset of convection outside the tangent cylinder flow structure appears as a equatorial 
symmetric. Though by increasing Rayleigh numbers the equatorial symmetric constrain 
impose by rotation break down. So for that the simulations are done in both low and high 
Rayleigh numbers. \par
In low-Ra regime structure of the flow is like an onset of convection. 
In this case we see that in both
nonlinear and kinematic case the dipole structure is preferred and initial
magnetic field structure belongs to quadrupole family becomes a fail dynamo. 
The dipolar magnetic field has an equatorial antisymmetric structure and quadrupolar
is equatorial symmtric. As the initial input poloidal magnetic field is very weak, 
the initial ampilification of the magnetic field in the nonlinear simulation 
is in kinematic regime. In both kinematic and nonlinear dynamos, the fall of the
magnetic energy of the quadrupole is effectively a kinematic process since the growth
of the field is not much to effect the flow at all. For initial dipole seed field start, 
both kinematic and nonlinear shows a dipolar strcuture. The analyses suggests that
as for quadrupole field, the $B_\phi$ peaks at the equatorial plane, hence
the columnar strcuture not able to reinforce any poloidal field to the 
original field as classical-$\alpha$ effect suggests and eventually dynamo fail due to 
incompleteness of the regeneration cylcle. 
While the columnar structure able to generate poloidal field from the toroidal field, 
hence complete the regeneration cycle in the kinematic regime. \par
Further to test role of the initial magnetic field in polarity selection, we
use a mixed seed field compose of majorly axial quadrupole and very little trace of 
axial dipole (99.99\% of $\Lambda_{20}^P$ and 0.01\% of $\Lambda_{10}^P$). In this 
case the final structure of the magnetic field shows up a dipolar strcucture. 
From begining of the simulation, we see that
the axial quadrupole is falling and axial dipole is growing. These results suggests that 
structure of the initial magnetic field is playing role in polarity selection. 
\par
Since, the input structure of the velocity field is equatorial symmetric 
under rapid rotation. Therfore to study the role of symmetry of the velocity field 
in polarity selection, we introduce seed
equatorial antisymmetric flow in the low-$Ra$ regime. The results suggest that
a successful dipolar dynamo in both kinematic and nonlinear regime from an axial
quadrupole seed magnetic field. This happens in the induction equation due to
interaction of the equatorial antisymmetric flow with the equatorial symmetric field. 
Though in high-$Ra$ case, where velocity symmetry break down; there we see
a dipole structure of the magnetic field in nonlinear dynamo. Though the kinematic model 
in this case shows a chaotic solution. Nevertheless, our results suggests
that dipole is preferred over quadrupole in all cases of rapidly rotating dynamos.
Therefore the preference of the dipole structure in the low-$Ra$ regime is 
a kinematic effect.  
\par
Our second part of the study is preferences of dipole in rapidly rotating dynamo 
models at high-$Ra$. 
Here, emphasise is on to quantify role of the magnetic field in dipole preferences
at different strength of the self generated magnetic field by varying Rayleigh numbers.
Our focus is on evolution of the field from a axial dipole seed magnetic field in
both nonlinear and kinematic regimes. 
\par
We look into the $\bm{B}^T_{20}$ generation for two different Rayleigh numbers. 
Energy analysis suggests that the field is generated by the shearing of axial 
dipole by differential rotation as the classical $\omega$-effect suggests. 
Energy matrices construct based on Bullard's selection rule
give an idea of the interaction between different harmonics of the flow and field to 
generate the field. For $Ra=140$, we see it is the $Y^0_1$ of the poloidal field 
is sheard by $Y^0_1$ of the toroidal field
in both kinematic and nonlinear regime. Though in $Ra=220$, the zonal flow is dominated 
by a $Y^0_3$ and energy matrix suggest that in the presence of strong dipole field, 
the differential rotation which is helping in kinematic regime switch from $Y^0_1$ 
to $Y^0_3$. Nevertheless, the structure of the
differentail rotation is still a large scale as classical $\omega$-effect shows. We do 
find that the $\omega$-effect is a robust features of a rapidly rotating dynamo. 
\par
In the last part of the study we discuss about generation of axial dipole field  
in kinematic and nonlinear dynamos. In high-$Ra$ we see effect of the back reaction due to
magnetic field by resorting back the dipole structute from a transient chaotic solution
before the saturation. Though kinematic dynamo does not show any such behaviour
as there is no effect of back reaction on the flow. The important results that has
come out from our study is the magnetic field acts on the flow much before the saturation.
Our study suggests that the growth of the magnetic field is not a kinematic effect as
one might think off, rather a dynamic effect. This dynamic effect grows as the field generated
in time and finally brings the saturation. Though we have not study the mechanism of the
saturation in this nonlinear models. Strength of the magnetic field in a self sustained dynamos
can be done by only varying Rayleigh number (energy input), unlike in magnetoconvection problem.    
In our model we vary Rayleigh number to see the effect of back reaction on the dynamics.
The results suggest that effect of back reaction delays as we decrease the Rayleigh number.
In our high-$Ra$ simulations, we find two regimes - the regime before departure between nonlinear 
and kinematic simulations as kinmetic and after (before the saturation) as nonlinear phases. 
In kinematic phase, we find that classical Parker dynamo model works quite well, though it does 
not help much in large scale magnetic field (axial dipole) generation. As the model evolves into 
a nonlinear phase, we do see a sudden growth of axial dipole. Energy and induction terms analyses 
provide a hint on a different generation mechanism. We find that the term which support the axial 
dipole in kinematic phase is becoming a negative contributer in the nonlinear phase. The detail
analyses suggest that the $r$-component of the axial dipole is supported by a advection of
gradient of small scale radial magnetic field. Though advection does not contribute to the 
overall magnetic energy amplification. Transport coefficients of mean field suggests that 
advection plays a important role in dipole field generation \cite{Schrinner2007} \cite{Schrinner2012}.

\clearpage  

\bibliography{draft}

\begin{appendices}
\numberwithin{equation}{section}
\section{Bullard's selection rule} \label{S}
Eq. \ref{eq5} can written as 
\begin{equation}\label{eqA1}
\begin{aligned}
\bm{u} = \sum\limits_{m=0}^{M} \sum\limits_{l=1}^{L} (\bm{u}_{lm}^{Tc} + \bm{u}_{lm}^{Ts} + \bm{u}_{lm}^{Pc} + 
\bm{u}_{lm}^{Ps}) 
\end{aligned}
\end{equation}
where, the vector harmonics are given as, for example
\begin{eqnarray}\label{eqA2}
\begin{aligned}
\bm{u}_{lm}^{Tc} = \nabla \times \nabla \times [u^{Tc}_{lm} (r) Y_l^m(\theta) \cos\phi \hat{\bm{r}}] \\ 
\bm{u}_{lm}^{Ts} = \nabla \times \nabla \times [u^{Ts}_{lm} (r) Y_l^m(\theta) \sin\phi \hat{\bm{r}}]
\end{aligned}
\end{eqnarray}
Similar way, we can expand the $\bm{B}$.

\subsection{Rules for \texorpdfstring{$\bm{B}^P_{10}$}{TEXT} and 
\texorpdfstring{$\bm{B}^P_{20}$}{TEXT}  generation} \label{S1}

\subsubsection{For combinations of \texorpdfstring{($\bm{u}^P$$\bm{B}^P$$\bm{B}^P$)} 
{TEXT}} \label{S11}  

The first two terms inside bracket denotes flow and magnetic field of the induction term
and last term represents the induced field. Superscripts $P$ and $T$ denote toroidal and
poloidal componenents of the vector field, respectively. Degree of 
the vectors are defined as $l_\alpha$, $l_\beta$ and $l_\gamma$ and
orders are as  $m_\alpha$, $m_\beta$ and $m_\gamma$ according to their appearance inside
the bracket. Note that our notation of degrees are different from the 
Bullard's notation \cite{Bullard1954}. \par
 
$(a)$ Rules for $m_\alpha$ and $m_\beta$ \par  According to
the selection rule for generating axisymmetric field both inducing flow and field will share
common value of order i.e., $m_\alpha=m_\beta$. This rule is valid for
any toroidal and poloidal combinations of flow and field. \par

$(b)$ Rules for $l_\alpha$ and $l_\beta$ \par
The restriction for degrees are 
$l_\alpha+l_\beta+l_\gamma$ is even and they can form the sides of a triangle 
(degenerate case are $l_\alpha = l_\beta+l_\gamma$, etc ) means
$|l_\alpha-l_\gamma| \leqslant l_\beta \leqslant (l_\alpha+l_\gamma) $.

Therefore, for generating $\bm{B}^P_{10}$ i.e., by fixing $(l_\gamma=1)$, range of
$l_\beta$ is between $(l_\alpha-1)$ and  $(l_\alpha+1)$. So the combinations are
$[1, 2]$, $[2, 1]$, $[2, 3]$, $[3, 2]$, $[3, 4]$, etc, where
first digit shows $l_\alpha$ and second shows $l_\beta$.    \par

Similarly, for generating $\bm{B}^P_{20}$ i.e., by fixing $(l_\gamma=2)$, the possible combinations of 
$l_\alpha$ and $l_\beta$ are $[1, 1]$, $[1, 3]$, $[2, 2]$, $[2, 4]$, etc.    

\subsubsection{For combinations of \texorpdfstring {($\bm{u}^P$$\bm{B}^T$$\bm{B}^P$)} {TEXT} and 
\texorpdfstring{($\bm{u}^T$$\bm{B}^P$$\bm{B}^P$)}{TEXT} } \label{S12}

$(a)$ Rules for $m_\alpha$ and $m_\beta$ \par
Rule of \ref{S11} $(a)$ holds for this combinations.   \par
$(b)$ Rules for $l_\alpha$ and $l_\beta$ \par
The range of $l$ are same as \ref{S11} $(b)$ except, 
$l_\alpha+l_\beta+l_\gamma$ is odd and two harmonics should not be identical.
The latter suggests that for example, for a particular choices of
combinations of $\bm{u}^{Tc}_{lm}$ and $\bm{B}^{Tc}_{lm}$, 
or $\bm{u}^{Ps}_{lm}$ and $\bm{B}^{Ps}_{lm}$ 
can not generate any poloidal magnetic field.   
So for generating $\bm{B}^P_{10}$, the combinations of $l_\alpha$ and
$l_\beta$ are $[1,1]$, $[2,2]$, $[3,3]$, $[4,4]$, etc (Note that, two harmonics 
should not be identical but they can share same $l$ and $m$). The combinations of $l_\alpha$ and
$l_\beta$ for $\bm{B}^P_{20}$ are $[1,2]$, $[1,3]$, $[2,1]$, $[2,3]$, etc. 
   
\subsection{Rules for \texorpdfstring {$\bm{B}^T_{20}$}{TEXT} generation} \label{S2} 
 
\subsubsection{For combinations of \texorpdfstring {($\bm{u}^P$$\bm{B}^T$$\bm{B}^T$)}{TEXT} and 
\texorpdfstring {($\bm{u}^T$$\bm{B}^P$$\bm{B}^T$)}{TEXT}}  \label{S21}

$(a)$ Rules for $m_\alpha$ and $m_\beta$ \par
Rule of \ref{S11} $(a)$ holds for this combinations.   \par
$(b)$ Rules for $l_\alpha$ and $l_\beta$ \par
The range of $l$s are same as \ref{S11} $(b)$. The combinations of $l_\alpha$ and
$l_\beta$ for $\bm{B}^T_{20}$ are $[1,1]$, $[1,3]$, $[2,2]$, $[2,4]$, etc.

\subsubsection{For combinations of \texorpdfstring{($\bm{u}^P$$\bm{B}^P$$\bm{B}^T$)}{TEXT} and 
\texorpdfstring {($\bm{u}^T$$\bm{B}^T$$\bm{B}^T$)}{TEXT}}  \label{S22}

$(a)$ Rules for $m_\alpha$ and $m_\beta$ \par
Rule of \ref{S11} $(a)$ holds for this combinations.   \par
$(b)$ Rules for $l_\alpha$ and $l_\beta$ \par
The range of $l$s are same as \ref{S12} $(b)$. The combinations of $l_\alpha$ and
$l_\beta$ for $\bm{B}^T_{20}$ are $[1,2]$, $[2,1]$, $[2,3]$, $[2,3]$, etc.

\end{appendices}

\end{document}